\documentclass[prb,twocolumn,showpacs,superscriptaddress,preprintnumbers,amssymb]{revtex4-2}
\usepackage{graphicx}% Include figure files
\usepackage{color}
\usepackage{amsmath}
\usepackage{dcolumn}% Align table columns on decimal point
\usepackage{bm}
\usepackage{slashed}
\usepackage{graphicx}
\usepackage{slashed}
\usepackage[normalem]{ulem}
\usepackage{comment,cancel}

\newcommand{\beq}{\begin{equation}}
\newcommand{\eeq}{\end{equation}}
\newcommand{\beqn}{\begin{eqnarray}}
\newcommand{\eeqn}{\end{eqnarray}}

\newcommand{\ra}{\rightarrow}

\newcommand{\cC}{ {\cal C} }

\newcommand{\cG}{ {\cal G} }

\newcommand{\cT}{ {\cal T} }
\newcommand{\cS}{ {\cal S} }

\newcommand{\cZ}{ {\cal Z} }
\newcommand{\tr}{\mathrm{tr}}

\newcommand{\ii}{\mathrm{i}}

\newcommand{\SU}{\mathrm{SU}}
\newcommand{\U}{\mathrm{U}}

\newcommand{\cxa}[1]{{\color{black} #1}}

\definecolor{orange_custom}{rgb}{0.93, 0.47, 0.2}

\begin{document}

\title{A Guide to Symmetric Mass Generation in Lattice-QCD}

\author{Anna Hasenfratz}

\affiliation{Department of Physics, University of Colorado, Boulder, Colorado 80309, USA}

\author{Cenke Xu}

\affiliation{Department of Physics, University of California, Santa Barbara, CA 93106, USA}

\date{\today}

\begin{abstract}

Symmetric mass generation (SMG) has attracted growing interest in both condensed matter theory and lattice-QCD communities. Here we formulate general criteria for SMG and examine their compatibility with lattice-QCD. We propose possible RG-flow scenarios near the SMG transition, and argue that meson mass ratio can serve as a probe of the SMG transition viewed as a UV fixed point. We further identify Goldstone tetraquark meson states as phenomenological signatures of the ``type-II'' SMG phase.

\end{abstract}

\maketitle

\section{Introduction}

The fate of strongly interacting fermions is a very challenging problem in both high energy and condensed matter physics. In the conventional scenario, strong interaction leads to the condensation of a quadratic-fermion (fermion bilinear) operator, i.e. the expectation value $\langle \bar{\psi} T \psi \rangle$ is nonzero, where $T$ is a Hermitian matrix in the flavor space. In this condensate, the fermions are massive (gapped), and the low-energy spectrum of the system is dominated by massless Goldstone bosons. These modes arise as a consequence of spontaneous symmetry breaking induced by the formation of the condensate. Familiar examples of this mass-generation mechanism include chiral and parity symmetry breaking in the confined phase of Quantum Chromodynamics (QCD). In this phase, quadratic-quark condensation spontaneously breaks chiral symmetry and simultaneously produces a spectral gap for quarks. The condensate also leads to the formation of light pseudo-Goldstone bosons, known as pions. In condensed matter systems, the most prominent example of this mass generation is the BCS theory, which is the mechanism of superconductivity in interacting electrons or neutral atoms such as $^3$He. There, the condensation of fermion-pairs spontaneously breaks a particle-number $\U(1)$ symmetry, leading to a gapped fermionic spectrum. Phase transitions driven by such symmetry-breaking mass generation can be described by the Gross-Neveu-Yukawa field theory, including its non-relativistic generalization.  

However, a new mechanism of mass generation without a quadratic-fermion condensate has attracted growing attention in the past decade. This phenomenon is now known as symmetric mass generation (SMG)~\cite{fidkowski1,fidkowski2,3450,Slagle,shailesh,shailesh2,shailesh3,catterall_2016,catterall2,catterall3,catterall4,mengSMG1,annaSMG2,SMGDQCP,SMGDQCP2,Tong_2021,Tong_2022,34502,disorderSMG,SMGreview}. Examples of SMG have been discussed analytically in (1+1)D~\cite{fidkowski1,fidkowski2,3450}, and also reported numerically in models across all dimensions below and equal to $(3+1)$~\cite{Slagle,shailesh,shailesh2,shailesh3,catterall_2016,catterall2,catterall3,catterall4,mengSMG1,disorderSMG,annaSMG1,annaSMG2,marina1,marina2}. Many previous works have also suggested the potential role of SMG in the lattice realization of the Standard Model and Grand Unified Theory~\cite{wenSM,xu16,xuGUT,Wang_2020}. Despite this progress, a systematic discussion of the conditions and the phenomenological consequences of SMG, especially its implications in lattice-QCD, is still lacking. It is the goal of the current work to bridge this gap in the literature. 

%We will use the staggered fermion realization of (3+1)D QCD as the potential platform of SMG in this work. 
We begin with a general discussion of the definition and necessary conditions of the SMG, and introduce two possible types of SMG. %Then we will show that the continuum Symanzik effective action of certain staggered fermions~\cite{leesharpe} does meet the key conditions for SMG. 
%Guided by the numerical data on staggered fermions, 
Assuming SMG is realized in certain lattice-QCD models, we propose possible RG flow around the SMG phase, and identify the meson mass ratio as a numerical probe for the SMG transition, and that the Goldstone tetraquark meson states provide a phenomenological signature of the ``type-II" SMG phase.

%In the next section we provide a general discussion of SMG.

\section{General discussion \label{sec:gendis}}

\subsection{Definition and Remarks}

The phenomenon of symmetric mass generation(SMG) can be defined as follows:

\medskip

%%(including both self-anomalies and mixed 't Hooft anomalies between $\cG$ and other exact global symmetries),
%\sout{If $\cG$ has no 't Hooft anomaly, the system can in principle be driven into a gapped and nondegenerate state by interaction while preserving $\cG$.} 

Consider a system whose noninteracting limit contains gapless fermions $\psi$. The system has a {\it key symmetry} $\mathcal{G}$, which is the (smallest) symmetry under which {\it all} fermion-bilinear mass operators including $\bar{\psi} T \psi$ and $\bar{\psi} \gamma_5 T \psi$, as well as the Majorana mass operators transform nontrivially. 
If $\cG$ is free of 't Hooft anomalies then, at least in principle, interactions can drive the system into a phase where the fermion $\psi$ is gapped while still preserving $\cG$.
In such a phase, the expectation value of any mass operator of $\psi$ remains zero, even though $\psi$ acquires a mass. 

We stress that anomaly cancellation is a {\it necessary}, but {\it not sufficient} condition for SMG. The anomaly analysis gives a kinematic constraint: it determines whether a fully symmetric, trivially gapped phase is allowed in principle for a class of models. Whether such an SMG phase is actually generated in a model with a specific set of parameters is a dynamical question, which depends on the detailed form of the UV action. %Thus, if the 't Hooft anomaly of $\cG$ is not cancelled, SMG is impossible; but even when the anomaly is cancelled, a particular model need not realize SMG without further deformation. 
In this guide we use staggered fermions as a potential platform for SMG, because they realize a natural symmetry $\cG$ in a lattice-regularized setting. Nevertheless, realizing the desired SMG phase may require additional UV deformations of the standard staggered-fermion lattice model, such as extra four-fermion interactions, while preserving the key symmetry $\cG$.

\medskip

There exist several possible choices for the key symmetry $\cG$ that can give rise to SMG, including Spin-$Z_4$ in (1+1)D and (3+1)D, as well as time-reversal $\cT$ (more precisely $\cC\cT$) in (0+1)D and (2+1)D. The Spin-$Z_4$ symmetry plays a central role in our discussion, as it can be naturally implemented in various lattice models as a non-onsite symmetry. In the infrared the Spin-$Z_4$ transformation is defined as
\beqn \psi \ra e^{\ii k \frac{\pi}{2}\gamma_5} \psi, \ \ k = 0, 1, 2, 3. 
\eeqn 
When combined with $\U(1)_V$, the Spin-$Z_4$ symmetry becomes equivalent to a chiral-$Z_2$ symmetry, under which left-handed Weyl fermions pick up a minus sign while right-handed Weyl fermions remain invariant. The self-anomaly classification shown in table~\ref{table} is defined with the convention that the minimal nontrivial element of the anomaly class is realized by a single complex Dirac fermion in the corresponding dimension. As an illustration, the Spin-$Z_4$ self 't Hooft anomaly classification in (3+1)D is $\mathbb{Z}_8$ in this table~\cite{spinZ41,spinZ42,spinZ43,xu16}, which implies that eight copies of Dirac fermions in (3+1)D (counting both flavors and colors) are required to cancel the associated ’t Hooft anomaly. 
%If instead one chooses \an{\sout{a} to work with}  Weyl fermions, then anomaly cancelation demands a total number of Weyl fermions that is a multiple of 16.
If, instead, one chooses to work with with Weyl fermions, then the requirement of anomaly cancellation implies that the total number of Weyl fermions must be an integer multiple of 16.

\begin{table}[h!]
\centering
\begin{tabular}{|c||c|c|c|c|}
\hline
 & & & & \\
dimension & (0+1) & (1+1) & (2+1) & (3+1) \\ & & & & \\
\hline
  & & & &\\
$\cG$ & $\cT, \ \cT^2 = 1$ & Spin-$Z_4$, & $\cT, \ \cT^2 = -1$ & Spin-$Z_4$ \\
  & & &  &\\
\hline
  & & & &\\
Anomaly class & $\mathbb{Z}_4$ & $\mathbb{Z}_4$, & $\mathbb{Z}_{8}$ & $\mathbb{Z}_{8}$ \\
  & & &  &\\
\hline

\end{tabular}

\caption{Examples of the ``key symmetry" $\cG$ in different dimensions that could lead to SMG. Here the anomaly class assumes the elementary state is a complex Dirac fermion of corresponding dimensions. For example, the Spin-$Z_4$ symmetry has a $\mathbb{Z}_8$ classification, when the elementary state is considered to be one complex Dirac fermion.} \label{table}
\end{table}

\medskip

{\bf Remark 1.} $\cG$ could be either an ultraviolet (UV) exact symmetry or an infrared (IR) emergent symmetry of the system. If $\cG$ is enforced, all quadratic-fermion mass terms are excluded; the corresponding mass operators must have zero expectation value.

\medskip

{\bf Remark 2.} The ``system" may refer to either a $D$-dimensional bulk or the $D$-dimensional boundary or domain wall of a $(D+1)$-dimensional bulk. $\cG$ is often realized as a non-onsite symmetry if the system itself is a bulk, but it could be an exact onsite symmetry~\footnote{An onsite symmetry means that the symmetry operator is the product of operators acting on local tensor-product Hilbert spaces; it also implies that the symmetry can be promoted to a gauge symmetry. A non-onsite symmetry would generally involve lattice transformations, such as translation.} if the system is at the boundary/domain wall of a higher-dimensional bulk.

\medskip

{\bf Remark 3.} The system can also have a gapless phase, with a larger symmetry $\tilde{G}$ {\it in the infrared}, $\tilde{G} \supseteq \cG$. This larger symmetry $\tilde{G}$ often has 't Hooft anomaly.

\medskip

{\bf Remark 4.} In the SMG phase, the anomalous symmetry $\tilde{G}$  {\it must be} broken, but $\cG$ is still preserved, meaning that quantities that transform nontrivially under $\tilde{G}$ but trivially under $\cG$ must acquire a nonzero expectation value in the SMG phase.

This remark implies that, although the SMG phase enforces a vanishing expectation value for all quadratic fermion mass operators, it can nonetheless exhibit spontaneous symmetry breaking. In this context, the qualifier “{\it symmetric}” in “symmetric mass generation” refers specifically to the symmetry group $\cG$, rather than to $\tilde{G}$. Any SMG phase is required to break all symmetries that possess a ’t Hooft anomaly—such as the $\SU(N_f)_L \times \SU(N_f)_R$ flavor symmetry in $(3+1)$ dimensions—as well as any symmetry that exhibits a mixed ’t Hooft anomaly with $\cG$. In the SMG phase, however, the chiral flavor symmetry is broken via the condensation of a higher-order fermion operator, instead of a quadratic-fermion mass condensate. In other words, the SMG phase breaks the chiral flavor symmetry down to a {\it different subgroup} of $\tilde{G}$ than the one selected by a fermion bilinear mass condensate. In examples discussed in this manuscript, the chiral flavor symmetry group includes a discrete subgroup $\cG = $ Spin-$Z_4$, which should remain unbroken in the SMG phase but is broken in the conventional massive phase where a quadratic fermion mass term condenses.

\medskip

{\bf Remark 5.} The discussion here does not apply to cases when the system is driven into a gapped topological order, which does not always require a vanishing 't Hooft anomaly of $\cG$. From another perspective, a gapped topological order spontaneously breaks certain higher-form symmetries. Therefore, when higher-form symmetries are taken into account, topological orders are not fully symmetric. In contrast, the SMG phase defined here is symmetric even when higher-form symmetries are included.

\medskip

{\bf Remark 6.} The notion of 't Hooft anomalies is essential to the understanding of SMG.

\medskip

When a global symmetry $G$ has a 't Hooft anomaly, it means that although $G$ is {\bf not} gauged, attempting to gauge it would lead to a gauge anomaly. For this reason, a 't Hooft anomaly is often described as ``{\it obstruction to gauging}".

\medskip

Since in the original theory $G$ is not gauged, it describes a consistent field theory. However, the presence of a 't Hooft anomaly in a global symmetry imposes constraints on how that symmetry can be realized in any consistent ultraviolet (UV) regularization, such as in lattice gauge theories. A well-known example is the Nielsen-Ninomiya no-go theorem~\cite{nn1,nn2} and its generalizations to interacting systems (see for example Ref.~\cite{shamir1,shamir2,fidkowskixu}).

\subsection{Two types of SMG:}

%\sout{Discussions in this section}
The discussion above suggests that there are two possible types of SMGs:

\begin{itemize}

    \item {\bf Type-I:} None of the system's symmetries carry a ’t Hooft anomaly, and the SMG phase is fully symmetric, fully gapped, and the ground state is unique (nondegenerate).

    \item {\bf Type-II:} The system possesses an enlarged symmetry $\tilde{G}$ that carries a ’t Hooft anomaly, and in the SMG phase this symmetry $\tilde{G}$ is spontaneously broken, while the smaller symmetry $\cG$ remains intact. Consequently, no quadratic fermion mass operator acquires a condensate, but higher-order operators, such as quartic fermion operators, do condense.
    A concrete example of such an anomalous $\tilde{G}$ is provided by the continuous chiral flavor symmetries of QCD. After the higher-order fermion operator condenses, the unbroken residual symmetry must be anomaly-free. The spontaneous breaking of the continuous chiral flavor symmetry down to $\cG$ via a higher-order fermion condensate still produces Goldstone bosons in the SMG phase; however, these Goldstone modes are not ordinary mesons but instead take the form of “tetraquark” bound states.

\end{itemize}

\subsection{Symmetries and Anomalies of QCD}

%\sout{Here we discussed QCD} 
In this section, we consider (3+1)D QCD-like systems with $N_f$ flavors of Dirac fermions coupled with the $\SU(N_c)$ gauge field. Let us list the key symmetries of this theory, as well as their anomalies. We note that the faithful symmetry of the theory should be the combination of the following symmetries after quotienting their common shared subgroup, which is often discrete. 

\begin{itemize}

    \item $\SU(N_f)_L \times \SU(N_f)_R$:
    This is a continuous chiral flavor symmetry, generally with 't Hooft anomaly, and it is identified as the $\tilde{G}$ symmetry in the previous section. When we gauge the chiral flavor symmetry $\SU(N_f)_L$, the $\SU(N_f)_L$ gauge field will in general suffer from a gauge anomaly, except in the special case described below. Consequently, according to the definition given in the previous section, the ungauged chiral flavor symmetry possesses a ’t Hooft anomaly.

    One special scenario is when $N_f = 2$ and $N_c$ is an even integer. In this scenario, gauging the $\SU(2)_L$ flavor symmetry will not lead to gauge anomaly. The perturbative anomaly cancels out for the SU(2) Lie-algebra since $\tr\left( \sigma^a \{\sigma^b, \sigma^c\} \right) = 0$; The global Witten's anomaly for $\SU(2)_L$ gauge field also vanishes for even $N_c$. But there is a mixed anomaly between SU(2)$_{L,R}$ and $\U(1)_V$.

    \item $\U(1)_A$ and Spin-$Z_4$:

    Without the $\SU(N_c)$ gauge field, the $\U(1)_A$ symmetry is an actual symmetry of the field theory (even though $\U(1)_A$ has 't Hooft anomaly). 
    With the presence of the $\SU(N_c)$ gauge field, $\U(1)_A$ is broken due to the ABJ-anomaly: under $\U(1)_A$ rotation $\psi \ra \exp(\ii \alpha \gamma_5) \psi$, the partition function acquires a phase 
    \beqn \exp\left( \ii 2 \alpha N_f \int d^4x \, \frac{1}{8\pi^2} \, \tr\left( F \wedge F \right) \right). \eeqn
    However, the $Z_4$ subgroup of $\U(1)_A$, i.e. the Spin-$Z_4$ symmetry, is still preserved as long as $N_f$ is even. The Spin-$Z_4$ plays the role of $\cG$ discussed in the previous section. 
    
    \item We now turn to the vector-like symmetries of this system. Vector-like symmetries have no self-anomaly, but they can possess mixed ’t Hooft anomalies with $\U(1)_A$ and its discrete subgroups \cite{newanomaly,newanomaly1,newanomaly2}, for example Spin-$Z_4$, as well as with other discrete symmetries~\cite{zohar2017}. Physically this mixed anomaly manifests as the breaking of the Spin-$Z_4$, after faithfully gauging the vector-like flavor symmetries. 
    
    A particularly transparent instance of this phenomenon occurs when $N_c$ is an odd integer. In this case, there exists a mixed ’t Hooft anomaly between the $\mathrm{Spin}\text{-}Z_4$ symmetry and the vector-like flavor symmetry group $\SU(N_f)_V$.
    
    To reveal this mixed anomaly, one can gauge the $\SU(N_f)_V$, then the $\SU(N_f)_V$ gauge field would break the Spin-$Z_4$ symmetry through the simplest ABJ-anomaly. The existence of this mixed anomaly implies that, if the SMG phase happens, these vector-like flavor symmetries should also be broken. To accurately evaluate this mixed anomaly, one often needs to account for the common discrete subgroups shared among $\SU(N_c)$, $\SU(N_f)_V$, and $\U(1)_V$, as they lead to fractionalization of fluxes. The complete analysis of these mixed anomalies pertinent to SMG, as well as their connection to other restrictions such as the Lieb-Schultz-Mattis theorem, will be deferred to a future work~\cite{future}. 

\end{itemize}

When $N_c = 2$ the flavor symmetry is enlarged to $\SU(2N_f)$ in the field theory, thanks to the pseudo-real nature of SU(2). The $\SU(2N_f)$ symmetry likewise has 't Hooft anomaly. In this work, we will not discuss this enlarged flavor symmetry, as the lattice model (for example the staggered fermion) has a much lower symmetry in the UV.

Of course, a symmetry with 't Hooft anomaly cannot be realized as an onsite symmetry in the underlying lattice model in the same dimension. 
Therefore, there cannot be an onsite $\U(1)_A$ or continuous chiral flavor symmetries on the lattice, with or without the gauge field $\SU(N_c)$.
Consequently, no exact onsite \(\U(1)_A\) symmetry or continuous chiral flavor symmetries can be realized on the lattice, irrespective of the presence or absence of the \(\SU(N_c)\) gauge field.
Therefore these anomalous symmetries must be broken by higher order terms in the action. For example, such higher order terms are derived for the staggered fermion in Ref.~\cite{leesharpe}. The staggered fermion realization of QCD does have a UV non-onsite symmetry that corresponds to Spin-$Z_4$ in the continuum. 

%Based on Table~\ref{table}, the Spin-$Z_4$ is free of 't Hooft anomaly when $N_f N_c$ is multiple of 8, which is consistent with the staggered fermion numerics which demonstrate SMG in SU(3) QCD with $N_f = 8$, and SU(2) QCD with $N_f = 4$.

\section{Beyond anomaly cancellation}

In the previous section, we discussed two possible categories of SMG phases: the type-I SMG, which remains fully symmetric and gapped, and the type-II SMG, which spontaneously breaks an anomalous symmetry $\tilde{G}$ while keeping $\cG$ intact. One example of $\tilde{G}$ is the $\SU(N_f)_L \times \SU(N_f)_R$ symmetry, which must be broken even in the SMG phase.Thus, it remains possible for Goldstone modes to appear due to the spontaneous breaking of the (continuous) group $\tilde{G}$, while still not inducing any quadratic-fermion mass condensate. When $\cG$ is Spin-$Z_4$, the most natural condensate that spontaneously breaks $\tilde{G}$ to $\cG$ is a quartic-fermion operator condensate.

However, the type-II SMG may be subject to extra constraints, such as the inequalities proven in Ref.~\cite{weingarten}, which demand that the lightest boson of the system be the pion, which is a pseudo-scalar fermion-bilinear bound state. Let us review the inequality proven in Ref.~\onlinecite{weingarten}. We start with the most basic version: \beq \cS_{\rm QCD} = \cS_A + \int d^4x \ 
\sum_{j = 1}^{N_f} 
\bar{\psi} \slashed{D}_A \psi. \label{S}\eeq $A$ is a dynamical $\SU(N_c)$ gauge field. $\slashed{D}_{A}$ is \beq \slashed{D}_A = \slashed{D}_{A,0} + m, \ \ \ \slashed{D}_{A,0} = \gamma_\mu (\partial_\mu - \ii \sum_{l} A^l_\mu t^l). \eeq where $t^l$ spans the $\SU(N_c)$ Lie Algebra. 
%$\delta_{ij}$ is an identity matrix in the flavor space, with $i,j = 1 \cdots N_f$. 
$\cS_A$ is the action of $A^l_\mu$, which can be taken as the Maxwell term: $\cS_A = \int d^4x \frac{1}{2g^2} \tr \left( F_{\mu\nu} F^{\mu\nu} \right)$. We note that although our goal is to study QCD with zero bare fermion mass, here we keep a small $m$ to avoid singularities from the zero modes of $\slashed{D}_A$. 

The partition function of the system is \beq \cZ = \int DA \left({\rm det}\slashed{D}_A \right)^{N_f} \exp\left( - \cS_A \right). \eeq The operator $\slashed{D}_{A,0}$ is anti-hermitian: \beqn \slashed{D}_{A,0} = - \slashed{D}_{A,0}^\dagger, \label{keyeq1} \eeqn which ensures that all eigenvalues of $\slashed{D}_{A,0}$ are imaginary;  $\slashed{D}_A$ also satisfies
\beqn \gamma_5 \slashed{D}_{A} \gamma_5 = - \slashed{D}_{A,0} + m = \slashed{D}_A^\dagger. \label{keyeq2} \eeqn Eq.~\ref{keyeq1},\ref{keyeq2} imply that all complex eigenvalues of $\slashed{D}_A$ are ``paired up", i.e. if $\ii \lambda + m$ is an eigenvalue, so is $- \ii \lambda + m$. Therefore $({\rm det} \slashed{D}_A)^{N_f}$ is real and non-negative for any even integer $N_f$ \footnote{In fact, if $\slashed{D}_{A,0}$ has no zero mode, ${\rm det} \slashed{D}_A$ is already positive. If we take into account of zero modes of $\slashed{D}_{A,0}$, $({\rm det} \slashed{D}_A)^{N_f}$ also includes a factor $m^{N_f n_0}$, where $n_0$ is the number of zero modes of $\slashed{D}_{A,0}$. Therefore an even integer $N_f$ ensures that $({\rm det} \slashed{D}_A)^{N_f}$ being positive for any nonzero $m$. }. This is the condition for many other powerful inequalities proven in QCD-like theories~\cite{witteninequality,vafawitten,vafawitten2}.

Now let us evaluate the correlation function of the chiral current operator $\bar{\psi} \gamma_5 \gamma_\mu T^a \psi$. Here $T^a$ is a Hermitian matrix that belongs to the $\SU(N_f)$ Lie-algebra. \beqn && | \langle (\bar{\psi} \gamma_5 \gamma_\mu T^a \psi)_{x}  (\bar{\psi} \gamma_5\gamma_\mu T^a \psi)_{x'} \rangle | \cr\cr & = & | \int DA \, {\cal M}[A] \, \tr \left( \gamma_5 \gamma_\mu T^a S^A_{x, x'} \gamma_5 \gamma_\mu T^a S^{A}_{x', x} \right) | \cr\cr & \leq & \int DA \, {\cal M}[A] \, |\tr \left( \gamma_5 \gamma_\mu T^a S^A_{x, x'} \gamma_5 \gamma_\mu T^a S^{A}_{x', x} \right)| \cr\cr &\leq& \int DA {\cal M}[A] \cr\cr && \sqrt{ \tr \left( T^a S^{A}_{x,x'} T^a S^{A \dagger}_{x, x'} \right) \tr \left( T^a S^{A}_{x',x} T^a S^{A \dagger}_{x', x} \right)} \cr\cr &=&  \int DA \, {\cal M}[A] \, \tr \left( \gamma_5 T^a S^{A}_{x,x'} \gamma_5 T^a S^{A}_{x', x} \right) \cr\cr &=& | \langle (\bar{\psi} \gamma_5 T^a \psi)_{x}  (\bar{\psi} \gamma_5 T^a \psi)_{x'} \rangle  |, \eeqn where $S^A_{x,x'} = (\slashed{D}_A^{-1})_{x,x'}$, and integral measure $ {\cal M}[A] = \frac{1}{\cZ} e^{- \cS_A }  \left( {\rm det} \slashed{D}_A \right)^{N_f}$. We only evaluated the connected part of the correlation, because the disconnected diagrams (including diagrams where $x$ and $x'$ are connected through gauge field lines) vanish as they involve $\tr(T^a) = 0$. 

This inequality relies on the assumptions that the integral measure for the bosonic field configuration is positive. Under these conditions, the correlation between $\bar{\psi} \gamma_5 \gamma_\mu T^a \psi$ would be upper-bounded by the pion correlation, the pions $\bar{\psi}\gamma_5 T^a \psi$ always have the strongest connected correlation, or equivalently the smallest mass.

%This inequality is at odds with the type-II SMG discussed previously, when a quartic-fermion condensate spontaneously breaks the $\SU(N_f)_L \times \SU(N_f)_R$ symmetry to $\SU(N_f)_V$. Here we need to assume that $m$ in $D_A$ is very small, so that there is an approximate chiral flavor $\SU(N_f)_L \times \SU(N_f)_R$ symmetry. When the (approximate) chiral flavor symmetry is spontaneously broken, 
%the (pesudo)-Goldstone modes will lead to a gapless pole of the correlation of the chiral current operator $J_\mu^{A,a} = \bar{\psi} \gamma_5 \gamma_\mu T^a \psi$. This means that 
%the chiral current operator cannot have a short-range correlation within the length scale $1/m$, and in the limit of $m \ra 0$, the correlation length of the chiral current diverges. Then based on the inequality discussed above, pions also cannot have short-range correlation, incompatible with the phenomenon of SMG.

This inequality, if holds, is incompatible with the type-II SMG scenario~\cite{Kogan}, in which a quartic-fermion condensate spontaneously breaks $\mathrm{SU}(N_f)_L \times \mathrm{SU}(N_f)_R$ down to $\mathrm{SU}(N_f)_V$ while all pions remain short-ranged. We work with a small but nonzero $m$ in $\slashed{D}_A$, so that the theory has an approximate $\mathrm{SU}(N_f)_L \times \mathrm{SU}(N_f)_R$ symmetry, which is restored as $m \to 0$. If this (approximate) chiral symmetry is spontaneously broken, the chiral flavor current couples to the (pseudo)-Goldstone bosons, so the chiral current correlator has a correlation length that diverges as $m \to 0$. But the Weingarten inequality derived above bounds the chiral flavor current correlator in terms of the pion correlator, which then cannot remain short-ranged in the $m \ra 0$ limit. Hence, the type-II SMG is ruled out.

The type-II SMG is only possible when the inequality is invalidated. We can deform Eq.~\ref{S} with a perturbation, such as an imaginary term in $\cS_A$, which violates the positivity of the measure of bosonic fields. We can also turn on a four-fermion interacting term which can be included in the lattice model, and its effect could be captured by a Yukawa-type coupling in the continuum. As was noted in Ref.~\cite{vafawitten}, Yukawa interaction often invalidates some of the assumptions of the inequality. 
%\beq \cS = \cS_A + \int d^4x \ \sum_{j = 1}^{N_f} \bar{\psi}_j D_A \psi_j - \frac{u}{2} \left( \bar{\psi}\Gamma\psi \right)^2. \label{S2}\eeq
%Let's assume that $\Gamma$ is a Hermitian matrix, and introduce a Hubbard–Stratonovich field: 
%\beq \cS = \cS_A +  \int d^4x \ \sum_{j = 1}^{N_f} \bar{\psi}_j D_A \psi_j - \phi \left( \bar{\psi}\Gamma\psi \right) + \frac{1}{2u} \phi^2. \label{S3}\eeq Then one can still go through the evaluation of the inequality discussed above by considering Dirac operator \beqn D_{A,\phi} = \gamma_\mu (\partial_\mu - \ii \sum_l A^l_\mu t^l) - \phi \Gamma + m. \eeqn 
%The key assumptions of the inequality no longer hold when $u > 0$, and (for example) $\Gamma$ being one or all components of the $\gamma_\mu$ matrix. With this choice of $\Gamma$, Eq.~\ref{keyeq2} would be invalidated. 

The discussion in this section is of particular relevance to the potential realization of SMG using domain wall fermions. In the framework of domain wall fermion, the system of interest is realized at the (3+1)D domain wall of a (4+1)D bulk, therefore it is possible that the system of interest has the $\SU(N_f)_L \times \SU(N_f)_R$ flavor symmetry. Therefore, if we were to preserve the $\SU(N_f)_L \times \SU(N_f)_R$ flavor symmetry in the action, the type-II SMG is the only SMG that can occur in this system. However, in order to actually realize the type-II SMG, we must violate the inequality stated above. We note that Ref.~\onlinecite{Kogan} also showed that the Weingarten inequality, if it holds, excludes breaking the chiral flavor symmetry by condensing high-order fermion bound states.
%The discussion in this section is also similar to the proof of the absence of spontaneous breaking of continuous vector-like flavor symmetry in a class of theories~\cite{vafawitten}, and the proof therein also applies to scenarios of symmetry breaking through higher order operator condensate. 

\section{SMG in lattice systems \label{sec:staggered action}}

\subsection{Symmetry and Continuum action}

Now let us assume that a QCD-like theory is realized in certain (3+1)D or 4D lattice model. In some of the infrared fixed point (IRFP) there could be a large emergent symmetries $\tilde{G}$, such as the $\tilde{G} \sim \SU(N_f)_L \times \SU(N_f)_R$, but these symmetries will be broken by terms with higher dimensions. 
For example, the staggered fermion is a lattice realization of Dirac fermions coupled with gauge fields. One copy of staggered fermion corresponds to four-flavors of Dirac fermions in the IR.
%The full UV symmetry
%of the staggered is complicated. %Here we just discuss a few
%symmetry in the continuum with UV
%correspondence: %The Spin-$Z_4$ and $\U(1)_\epsilon$
%transformations are
%\beqn {\rm Spin-}Z_4 &:& \psi \ \ \ra \ \ e^{\ii k \frac{\pi}{2}
%\gamma_5 } \psi, \cr\cr \U(1)_{\epsilon} &:& \psi \ \ \ra \ \
%e^{\ii \theta \gamma_5 \xi_5} \psi, \cr\cr Z^\xi_{2} &:& \psi \ \
%\ra \ \ \xi_5 \psi. \eeqn
As was pointed out in Ref.~\onlinecite{leesharpe}, at finite lattice spacing the continuum staggered fermion action contains many extra terms that reflect the UV symmetries, which are lower than the IR symmetries $\tilde{G}$. For example, for one staggered fermion, i.e. $N_f = 4$, these extra terms include (but not limited to)~\cite{golterman}
\beqn \delta \cS = \int d^4x - \frac{u}{2}\left( (\bar{\psi}\psi)^2 - (\bar{\psi}\gamma_5 \xi_5 \psi)^2 \right) + \cdots.\, , \label{deltaS}\eeqn
where $\xi_5$ is a $4 \times 4$ matrix that acts in the flavor space. A more complete action 
%in the continuum
is given by Ref.~\onlinecite{leesharpe}. These extra terms break the anomalous $\SU(4)_L \times \SU(4)_R$ down to a much smaller vector-like flavor symmetry, and also break the anomalous $\U(1)_A$ down to Spin-$Z_4$ (as we discussed before, with a SU(2) gauge field, the global flavor symmetry of the QCD is enlarged to $\SU(8)$ thanks to the pseudo-real nature of SU(2). But here we focus on the $\SU(4)_L \times \SU(4)_R$ flavor symmetry given its explicit chiral nature). 
The remaining symmetries include a Spin-$Z_4$, a $\U(1)_\epsilon$, as well as a $Z_2^\xi$ symmetry: 
\beqn 
\text{Spin-}Z_4 &:& \psi \ \ \ra \ \ e^{\ii k \frac{\pi}{2} \gamma_5 } \psi, \cr\cr \U(1)_{\epsilon} &:& \psi \ \ \ra \ \ e^{\ii \theta \gamma_5 \xi_5} \psi. \cr\cr Z_2^\xi &:& \psi \ \ \ra \ \ \xi_5 \psi. 
\eeqn
$Z^\xi_2$ is a subgroup of $\Gamma_4$~\cite{leesharpe}, a discrete flavor symmetry generated by the product of four flavor $\Gamma$ matrices $\xi_\mu$. These are referred to as the “shift’’ symmetries of staggered fermions, which correspond to translations of the fermion field on the lattice. Consequently, the Spin-$Z_4$ symmetry is realized as a non-onsite symmetry in the staggered fermion formulation, emerging from a combination of $\U(1)_\epsilon$ and the shift symmetry. The anomaly classification Table~\ref{table} indicates that we need in total multiple of eight copies of Dirac fermions to ensure the cancellation of the self-anomaly of Spin-$Z_4$, meaning $N_f N_c$ must be a multiple of eight. This condition is satisfied by, for example, the $N_f = 4$ SU(2) lattice-QCD, and $N_f = 8$ SU(3) lattice-QCD.

As we explained earlier, anomaly cancellation is a necessary but not sufficient condition for realizing SMG. But if SMG is realized in certain lattice-QCD theories free of Spin-$Z_4$ self-anomaly, the additional four-fermion terms including those in Eq.~\ref{deltaS} would play a key role in understanding the SMG. These terms explicitly break the most obvious enlarged anomalous chiral flavor symmetry, thereby opening the possibility of a type-I SMG scenario. %Given the complexity of the four-fermion terms in Ref.~\onlinecite{leesharpe}, we leave a complete symmetry and anomaly analysis to future work. 
In principle One can also explore a  parameter space including extra UV deformations, provided these deformations are invariant under Spin-$Z_4$, which is the key symmetry $\cG$. If these higher dimension terms are absent, the continuum action would preserve a continuous chiral flavor symmetry, which would rule out type-I SMG and could at best allow a type-II SMG scenario. Such a scenario, however, is prohibited by the inequality discussed in the previous section.

\subsection{``Dimensional Reduction"}

We now analyze the possibility of SMG in lattice-QCD, following the logic of ``{\it dimensional reduction}" or ``{\it decorated defect}" arguments often used in the condensed matter literature (see for example Ref.~\cite{chenluashvin,senthilashvin,else,Li_2024,mawang}). The dimensional-reduction argument should be viewed as a heuristic mechanism rather than a rigorous derivation. We consider the example of the $N_f=4$ SU(2) QCD, with the $\delta \cS$ terms in Eq.~\ref{deltaS}.
We first introduce a Hubbard–Stratonovich field for $\delta \cS$: 
\beqn \delta \cS \sim \int d^4x - \phi_1 (\bar{\psi} \psi) - \phi_2 (\ii \bar{\psi}\gamma_5\xi_5 \psi) + \frac{1}{2u} |\vec{\phi}|^2. 
\eeqn
$\vec{\phi} = (\phi_1, \phi_2)$ is a two-component vector that rotates under $\U(1)_\epsilon$. The ``dimensional-reduction" argument starts with the ordered phase of $\vec{\phi}$. This phase has a nonzero condensate of $\vec{\phi}$, which spontaneously breaks $\U(1)_\epsilon$, and this ordered phase can be viewed as a ``$\U(1)_\epsilon$-superfluid" phase.

Our goal is to realize a fully gapped SMG phase, which requires driving the system into a disordered phase of $\U(1)_\epsilon$. Disordering a superfluid phase, one needs to proliferate the vortex loops of $\vec{\phi}$. A completely gapped and symmetric phase can only be achieved if these vortex loops do not trap any gapless modes. Fortunately, the modes trapped in the vortex loop can be solved directly. There are nonchiral (1+1)D fermions in the vortex loop, and the $\xi_5$ plays the role as $\tilde{\gamma}_5$ in this (1+1)d system: the right and left moving fermion modes carry $\xi_5 = \pm 1$ respectively. Therefore, along this vortex loop, the system reduces to a (1+1)D QCD with SU(2) gauge fields, and $\tilde{N}_f = 2$
\beqn \tilde{\cS} = \int d^2x \ \sum_{j = 1}^{\tilde{N}_f} \bar{\psi}_j \tilde{\slashed{D}}_{a} \psi_j + \cdots, \cr \cr \xi_5 \psi_{L} = \psi_L, \ \  \xi_5 \psi_R = - \psi_R. 
\eeqn
%Since $\xi_5$ is only an order-2
%symmetry (effectively a $Z_2$ symmetry) in staggered fermion, it
%is known that this (1+1)-dim system can be gapped through
%interaction without breaking any symmetry, as
The ellipses stand for extra terms, including the reduction of other dimension-6 four-fermion terms in Ref.~\onlinecite{leesharpe} into this vortex-loop (1+1)D space-time. We note that the four-fermion terms are much more relevant in this reduced (1+1)D space-time. 

Unlike $(3+1)$D, (1+1)D interacting fermion systems are much better understood, through various theoretical tools such as (nonabelian) bosonization~\cite{Witten1984}. A good starting point for understanding the $\tilde{N}_f = 2$ SU(2) QCD in (1+1)D is the coset decomposition of CFT: \beqn %\U(2\tilde{N}_)_1 = \SU(2)_{\tilde{N}_f} \oplus \Sp(2\tilde{N}_f)_1 . 
\U(4)_1 = \SU(2)_2 \oplus \left( \SU(2)_2 \oplus \U(1)_4 \right). \eeqn $\U(4)_1$ represents the free fermion CFT in (1+1)D, which can be decomposed into different degrees of freedom, as shown in the equation above. Gauging it with the $\SU(2)$ gauge field would erase the first $\SU(2)_{2}$ part of the decomposition, leaving only the $\SU(2)_2 \oplus \U(1)_4$ CFT as a coset~\cite{affleck1986,James_2018,delmastro2023}. 
%But this coset expression of CFT is unimportant for our central statement about the possibility of SMG phase along the vortex. 

We would like to explore the possibility of driving this (1+1)D CFT into a SMG phase. \cxa{The $Z_2^\xi$ symmetry acts just like the (1+1)D chiral-$Z_2$ symmetry along the vortex loop, and it is equivalent to Spin-$Z_4$ when combined with $\U(1)_V$. Chiral-$Z_2$ and Spin-$Z_4$  have classification $\mathbb{Z}_4$ in (1+1)D, according to Table~\ref{table}.} The chiral-$Z_2$ symmetry is free of anomaly for multiple of four Dirac fermions in (1+1)D, which is {\it exactly} the case here, as $\tilde{N}_f N_c = 4$ for this reduced (1+1)D system along the vortex loop. Therefore Table~\ref{table} suggests that the (1+1)D system can indeed be driven into a SMG phase. The $\SU(2)_2 \oplus \U(1)_4$ CFT is a critical point of a two-color and two-flavor interacting $1d$ quantum fermion system at half-filling, and the Spin-$Z_4$ is still implemented in the UV as lattice translation, the same as the shift symmetry of staggered fermion. In fact the $\SU(2)_2$ CFT is a well-known critical point next to the Haldane phase of spin-1 chain, which is a fully symmetric trivially gapped phase~\cite{affleck1986}. All these observations suggest that with proper four-fermion interactions, this vortex loop can be driven into the fully symmetric gapped SMG phase. Although we do not fully determine the form of the four-fermion interactions here, we know that they must preserve the Spin-$Z_4$, but meanwhile break the $\SU(2)_L \times \SU(2)_R$ chiral flavor symmetry of this (1+1)D system, to avoid 't Hooft anomaly. 

When the vortex loop is trivially gapped, one can in principle safely proliferate the gapped vortex loops, restore the $\U(1)_{\epsilon}$ symmetry in the (3+1)D, while keeping the entire system in a symmetric gapped phase, i.e. the SMG phase. More precisely, the final phase will preserve the $\U(1)_\epsilon$ and $Z_2^\xi$ symmetries, which together contain Spin-$Z_4$ as a subgroup. Here we made a nontrivial assumption that, away from the vortex loop, the gauge field is in its gapped confined phase, and its role along the vortex loop is just like a $(1+1)$D gauge field.  We note that physics of vortices in strongly interacting gauge theories can be studied in a much more reliable way in supersymmetric field theories~\cite{susyvortex1,susyvortex2,susyvortex3}. 

The same dimensional reduction argument also applies to other scenarios, for example $N_f = 8$ Dirac fermions coupled to SU(3) gauge field. Within the $\U(1)_\epsilon$ vortex loop, one effectively obtains $\tilde{N}_f = 4$ flavors of (1+1)D QCD with an SU(3) gauge field, again possessing the chiral-$Z_2$ symmetry. This theory likewise remains free from any Spin-$Z_4$ ’t Hooft anomaly, but we stress again that special care is required for the potential mixed anomaly between Spin-$Z_4$ and vector-like symmetries, which we defer to a future paper for systematic discussion~\cite{future}. 

%\begin{figure}
%    \centering
%    \includegraphics[width=\linewidth]{mass.pdf}
%    \caption{}
%    \label{mass}
%\end{figure}

\subsection{Possible critical behavior}

\begin{figure}
    \centering
\includegraphics[width=0.75\linewidth]{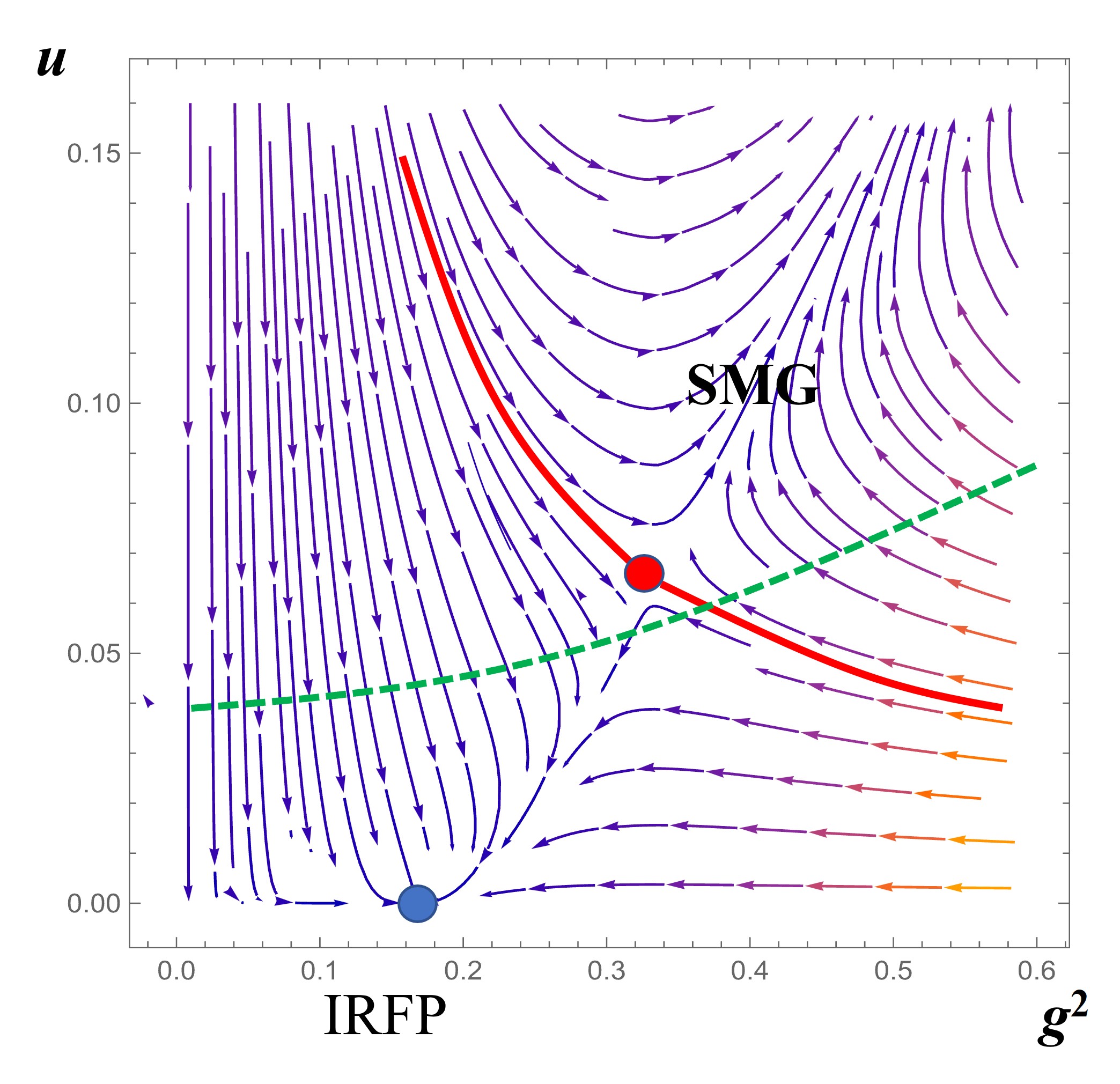}  
    \caption{ 
    Sketch of possible RG diagram for SMG. We consider cases in which $N_f N_c$ is a multiple of $8$, and the pair $(N_f,N_c)$ lies within the conformal window of the theory without $\delta \cS$. In the plot, $g^2$ denotes the gauge coupling, and $u$ parametrizes perturbations that explicitly break the anomalous chiral symmetry $\SU(N_f)_L \times \SU(N_f)_R$, including the term in Eq.~\ref{deltaS} and other higher-dimensional operators preserving Spin-$Z_4$. Since $(N_f,N_c)$ is assumed to lie within the conformal window, the theory at $u=0$ has an infrared fixed point with emergent $\SU(N_f)_L \times \SU(N_f)_R$ symmetry. In addition, there is a second fixed point at finite $u^\ast$, which is identified with the SMG critical point. The solid red line denotes the phase boundary separating the weak-coupling conformal phase from the strong-coupling SMG phase, while the dashed green line represents the trajectory corresponding to the lattice model. 
    }
    \label{RG}
\end{figure}

We now discuss possible critical behavior assuming that an SMG transition is realized in a (3+1)D or 4D lattice-QCD model, with a Spin-$Z_4$ symmetry and certain combination of $N_f$ and $N_c$ obeying the constraint $N_f N_c$ being multiple of eight. This theory could be staggered fermions plus certain UV deformations, such as extra higher dimensional terms in addition to the ones discussed in Ref.~\cite{leesharpe}. We also assume that $(N_f, N_c)$ lies within the conformal window, so the pure QCD theory (without extra higher-dimensional terms) has an IR conformal fixed point.  %For the Spin-$Z_4$ anomaly to be canceled, the product $N_f N_c$ must be a multiple of $8$ at all times. 
As previously emphasized, it is essential to retain the extra terms (which we assume has an overall scale $u$) breaking the anomalous $\tilde{G}$ in the analysis of the continuum action of SMG. %Both diagrams in Fig.~\ref{RG} apply to the case when the  $(N_f,N_c)$ system is within the conformal window, as there is an IR fixed point on the  $u = 0$ axis. 
There are two fixed points in the RG diagram sketched in Fig.~\ref{RG}: an IR fixed point with $u = 0$ and hence a large $\tilde{G} = \SU(N_f)_L \times \SU(N_f)_R$ symmetry, as well as a UVFP at $u^\ast \neq 0$ which we identify as the SMG transition. The IR symmetry at the UVFP is obviously lower than that at the IRFP. This means that states connected by $\tilde{G}$ should be degenerate at the IRFP, but the degeneracy is lifted at the UVFP. 
%The SMG transition in this case is a second order transition. 
%In both panels $(a)$ and $(b)$, a SMG phase transition can be driven by tuning the parameter in the lattice model, assuming the IRFP in panel $(b)$ is the stable attractive fixed point for a finite part of the diagram. These two scenarios can be discerned by investigating the symmetry at the SMG transition. 
For example, if $N_f = 4$, we can measure the mass ratio $R = M_S/M_{i5}$ of the following two mesons: \beqn  S: \bar{\psi} \left( {\bf 1} \otimes {\bf 1} \right) \psi \quad \text{and} \quad  \pi_{i5}: \bar{\psi} \left( \gamma_5 \otimes \xi_i \xi_5 \right) \psi, 
\eeqn
where $S$ is the scalar, and $\pi_{i5}$ is a pseudoscalar but has a flavor assignment $\xi_i\xi_5$. These two are onsite and ``1-link" operators in staggered fermion, easier to measure than mesons corresponding to higher-link operators.  
In the weak coupling phase $\beta > \beta_c$, $R$ should approach 1 in the thermodynamics limit, because $M_{\rm S}$ and $M_{i 5}$ are connected through the chiral rotation $\SU(4)_L \times \SU(4)_R$, an emergent symmetry of the IR conformal fixed point.  

\begin{figure}
    \centering    
    \includegraphics[width=0.75\linewidth]{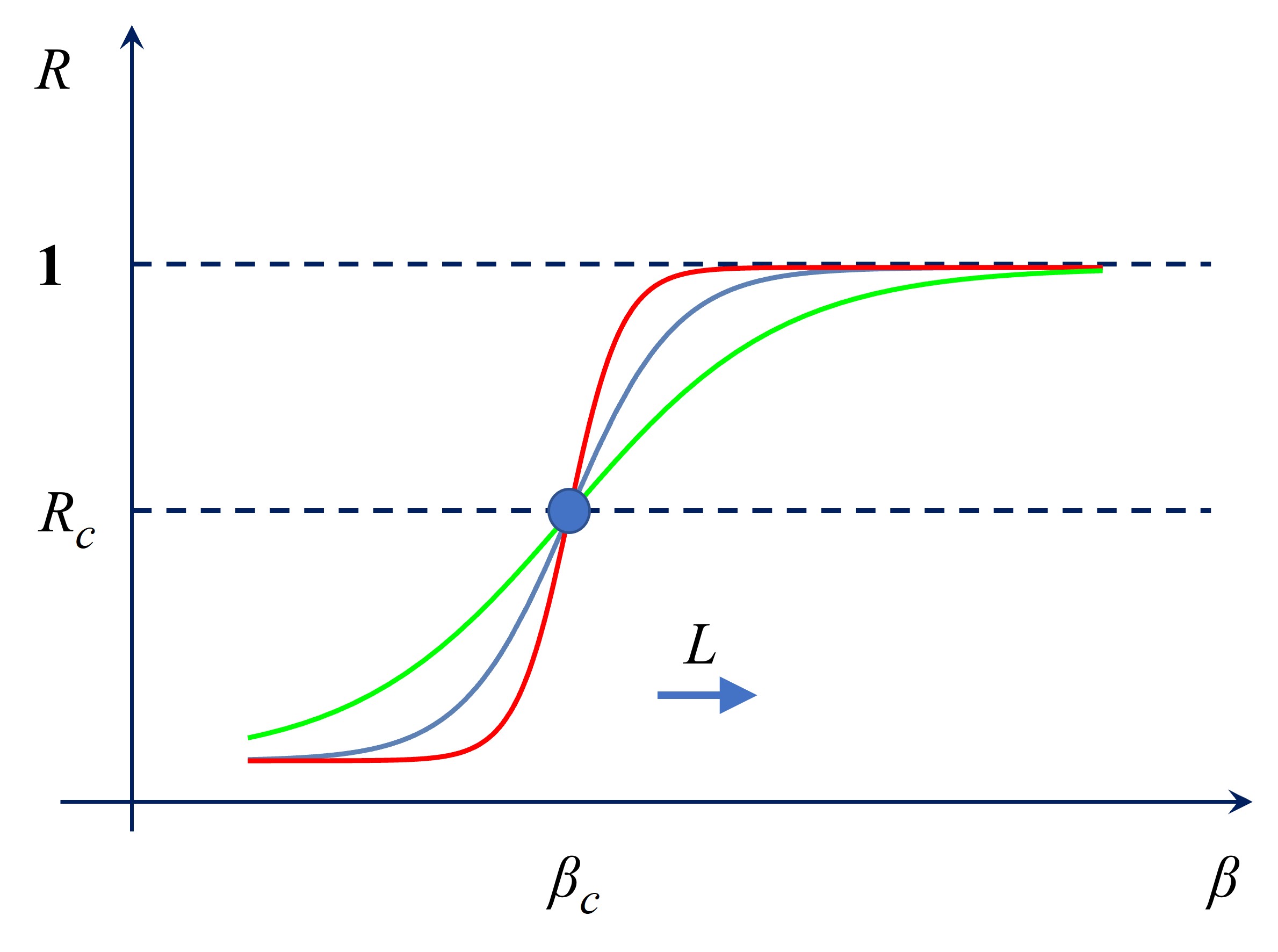}
    \caption{The predicted $R = M_{S}/M_{i5} $ based on RG diagram Fig.~\ref{RG}. In RG diagram Fig.~\ref{RG}, since the UVFP (the SMG transition) has lower symmetry than the IRFP, the mass ratio should differ from $1$ right at the transition $\beta = \beta_c$. The functions $R(\beta)_L$ with different $L$ are expected to cross at the same point $R_c$. }  
    \label{massratioplot}
\end{figure}

%The two scenarios of RG diagrams give different predictions for the mass ratio $R$ at the SMG transition $\beta_c$. 

The SMG transition corresponds to another fixed point with $u^\ast \neq 0$, suggesting that at the SMG transition there is no IR $\SU(4)_L \times \SU(4)_R$ symmetry, leading to a different value of the mass ratio $R_c \neq 1$. In fact, the RG diagram Fig.~\ref{RG} implies that in the thermodynamic limit $R$ would jump from $1$ for $\beta > \beta_c$, to another value $R_c$ at $\beta = \beta_c$. We stress that this discontinuous jump only occurs in the thermodynamic limit $L \ra \infty$. At any finite $L$ the ratio $R$ is a smooth function of $\beta$, and the functions $R(\beta)_L$ should cross at the same point $R_c$. This is because we expect the dimensionless number $R$ to be a universal function of $\beta - \beta_c$ and $L$: \beqn R(L^{1/\nu} (\beta - \beta_c)), \ \ \ R(+\infty) = 1. \eeqn $R_c = R(0)$ at $\beta = \beta_c$ is a universal value. The RG diagram Fig.~\ref{RG} predicts $R(\beta)_L$ as sketched in Fig.~\ref{massratioplot}.

This ``jump" of dimensionless quantities at a quantum phase transition is analogous to the behavior of the Binder cumulant at a second order phase transition. Similar behaviors of dimensionless quantities have been discussed in various contexts in condensed matter literature. %For example, for (2+1)-dim systems, the conductivity is a dimensionless value times $e^2/h$, and therefore it can take a universal value at a quantum critical point~\cite{fisher2}. It was then proposed that at the interaction-driven metal-insulator transition, the conductivity has a universal jump~\cite{senthilMIT,resistivity2}, similar to Fig.~\ref{massratioplot}$(a)$. 

By tuning $N_f$ and $N_c$, it is also possible that the UVFP and IRFP merge into one, and there is no UVFP with lower symmetry than IRFP. We note that $u$ is strongly irrelevant at the Gaussian fixed point, but with finite $g^2$ it may acquire a large anomalous dimension, and become relevant at the IR fixed point. In this case $u$ is only irrelevant at the Gaussian fixed point $g^2 = 0$, and it is often referred to as being ``dangerously irrelevant" in condensed matter. 

%When $\tilde{G}$ corresponds to the chiral flavor symmetry $\SU(N_f)_L \times \SU(N_f)_R$, 

\subsection{Goldstone modes of Type-II SMG}

As we discussed previously, if the model does have an anomalous symmetry $\tilde{G}$ such as the continuous chiral flavor symmetry, it must be spontaneously broken in the SMG phase by the condensation of higher-dimensional operator rather than fermion bilinears, which is what we called the type-II SMG. Such an order parameter may be constructed from four-fermion operators.

Despite the absence of a bilinear condensate, the spontaneous breaking of $\tilde{G}$ still gives rise to Goldstone modes. These appear as massless poles in correlation functions of the axial current,
\begin{equation}
J^{A,a}_\mu \sim \bar{\psi}\,\gamma_\mu\gamma_5\,T^a\,\psi,
\end{equation}
where $T^a$ are generators of $\SU(N_f)$. However, the spectrum differs qualitatively from that of conventional chiral symmetry breaking. All fermion-bilinear mass operators remain gapped, while the Goldstone modes are instead created by higher-dimensional operators.

In particular, if the order parameter is a four-fermion operator $P_4$, then its variation under a symmetry transformation,
\begin{equation}
\Pi^a \sim \ii\,[Q^a, P_4],
\end{equation}
is itself a four-fermion operator. This implies that the corresponding Goldstone bosons are created by local tetraquark operators. Consequently, the low-energy excitations are qualitatively different from conventional mesons. Furthermore, since the type-II SMG phase still has an unbroken Spin-$Z_4$, the Goldstone boson will not mix with any operator that is odd under Spin-$Z_4$, such as the fermion-bilinear mass operators.

\section{Discussion}

In this work, we discussed the general criteria and phenomenology of the symmetric mass generation (SMG) phase and the associated SMG transition. We distinguish between two qualitatively different realizations of SMG. In the type-I SMG scenario, all global symmetries are free of ’t~Hooft anomalies. As a result, the system can flow to a fully symmetric, fully gapped phase with a unique ground state, without the need for spontaneous symmetry breaking. We applied these ideas to lattice-QCD and argued that, if SMG is realized in these systems, higher-order operators in the Symanzik effective action would play a crucial role, as they explicitly break the obvious anomalous chiral flavor symmetry of the continuum theory, thereby opening the possibility of a type-I SMG phase. In contrast, the type-II SMG scenario arises when part of the symmetry, denoted $\tilde{G}$, carries a nontrivial ’t~Hooft anomaly. In this case, the SMG phase necessarily exhibits spontaneous breaking of $\tilde{G}$, but does so without generating a fermion bilinear condensate.

% Furthermore, we argue that recent numerical observations are consistent with the type-I SMG scenario within this framework. Motivated by these results, we have proposed a candidate renormalization-group flow for $N_f=4$ SU(2) gauge theory realized with staggered fermions.

The realization of type-II SMG is subject to additional nontrivial constraints. In particular, the theory must evade Weingarten-like inequalities that would otherwise forbid such a pattern of symmetry realization. Moreover, the unbroken symmetry group $\mathcal{G}$ in the SMG phase must be free of anomalies, including both its intrinsic ’t~Hooft anomalies and any mixed anomalies with other unbroken global symmetries. Assuming these conditions are satisfied, the resulting phase exhibits distinctive and unconventional phenomenology.

We would also like to clarify the current status of numerical search of SMG phases. Our latest numerical data suggests the existence of shift symmetry breaking in the strongly coupled phase in the 4D lattice-QCD model reported in Ref.~\cite{annaSMG2}, which also breaks $Z_2^\xi$ and Spin-$Z_4$. Similar shift symmetry breaking was also observed in the other 4D lattice-QCD model~\cite{annaSMG1}. Therefore, with these orders, the observed strongly coupled phase reported in these 4D models does not meet the most general definition of the SMG phase given in this work, although they are different from the standard confined phase of QCD as it preserves the $\U(1)_\epsilon$ symmetry. We will present the new numerical data and theoretical analysis in an upcoming paper. In contrast, in the $(2+1)$D SMG model~\cite{mengSMG1} without gauge field, the possible lattice-symmetry-breaking channels were explicitly examined but not observed in the numerical studies. 

\medskip

{\bf --- Acknowledgment}

\medskip

The authors thank Nouman Butt, Simon Catterall, Ho Tat Lam, Zohar Komargodski,
Roger Melko, Dan Sehayek, Yigal Shamir, Maarten Golterman, Mark Srednicki, David Tong, for very helpful discussions. Xu is supported by the Simons Foundation through the Simons Investigator program. Hasenfratz is supported in part by the U.S. Department of Energy, Office of Science, under Awards No.\ DE-SC0010005.  This research was supported in part by grant NSF PHY-2309135 to the Kavli Institute for Theoretical Physics (KITP), as the authors started this collaboration during KITP program ``What is Particle Theory?". 

During the preparation of this work the authors used GPT-5.4 Thinking to assist with literature search and to check parts of the derivations. After using this tool/service, the authors reviewed and edited the content as needed and take full responsibility for the content of the published article.

\bibliography{big}

\begin{thebibliography}{61}%
\makeatletter
\providecommand \@ifxundefined [1]{%
 \@ifx{#1\undefined}
}%
\providecommand \@ifnum [1]{%
 \ifnum #1\expandafter \@firstoftwo
 \else \expandafter \@secondoftwo
 \fi
}%
\providecommand \@ifx [1]{%
 \ifx #1\expandafter \@firstoftwo
 \else \expandafter \@secondoftwo
 \fi
}%
\providecommand \natexlab [1]{#1}%
\providecommand \enquote  [1]{``#1''}%
\providecommand \bibnamefont  [1]{#1}%
\providecommand \bibfnamefont [1]{#1}%
\providecommand \citenamefont [1]{#1}%
\providecommand \href@noop [0]{\@secondoftwo}%
\providecommand \href [0]{\begingroup \@sanitize@url \@href}%
\providecommand \@href[1]{\@@startlink{#1}\@@href}%
\providecommand \@@href[1]{\endgroup#1\@@endlink}%
\providecommand \@sanitize@url [0]{\catcode `\\12\catcode `\$12\catcode `\&12\catcode `\#12\catcode `\^12\catcode `\_12\catcode `\%12\relax}%
\providecommand \@@startlink[1]{}%
\providecommand \@@endlink[0]{}%
\providecommand \url  [0]{\begingroup\@sanitize@url \@url }%
\providecommand \@url [1]{\endgroup\@href {#1}{\urlprefix }}%
\providecommand \urlprefix  [0]{URL }%
\providecommand \Eprint [0]{\href }%
\providecommand \doibase [0]{https://doi.org/}%
\providecommand \selectlanguage [0]{\@gobble}%
\providecommand \bibinfo  [0]{\@secondoftwo}%
\providecommand \bibfield  [0]{\@secondoftwo}%
\providecommand \translation [1]{[#1]}%
\providecommand \BibitemOpen [0]{}%
\providecommand \bibitemStop [0]{}%
\providecommand \bibitemNoStop [0]{.\EOS\space}%
\providecommand \EOS [0]{\spacefactor3000\relax}%
\providecommand \BibitemShut  [1]{\csname bibitem#1\endcsname}%
\let\auto@bib@innerbib\@empty
%</preamble>
\bibitem [{\citenamefont {Fidkowski}\ and\ \citenamefont {Kitaev}(2010)}]{fidkowski1}%
  \BibitemOpen
  \bibfield  {author} {\bibinfo {author} {\bibfnamefont {L.}~\bibnamefont {Fidkowski}}\ and\ \bibinfo {author} {\bibfnamefont {A.}~\bibnamefont {Kitaev}},\ }\href@noop {} {\bibfield  {journal} {\bibinfo  {journal} {Phys. Rev. B}\ }\textbf {\bibinfo {volume} {81}},\ \bibinfo {pages} {134509} (\bibinfo {year} {2010})}\BibitemShut {NoStop}%
\bibitem [{\citenamefont {Fidkowski}\ and\ \citenamefont {Kitaev}(2011)}]{fidkowski2}%
  \BibitemOpen
  \bibfield  {author} {\bibinfo {author} {\bibfnamefont {L.}~\bibnamefont {Fidkowski}}\ and\ \bibinfo {author} {\bibfnamefont {A.}~\bibnamefont {Kitaev}},\ }\href@noop {} {\bibfield  {journal} {\bibinfo  {journal} {Phys. Rev. B}\ }\textbf {\bibinfo {volume} {83}},\ \bibinfo {pages} {075103} (\bibinfo {year} {2011})}\BibitemShut {NoStop}%
\bibitem [{\citenamefont {Wang}\ and\ \citenamefont {Wen}(2023)}]{3450}%
  \BibitemOpen
  \bibfield  {author} {\bibinfo {author} {\bibfnamefont {J.}~\bibnamefont {Wang}}\ and\ \bibinfo {author} {\bibfnamefont {X.-G.}\ \bibnamefont {Wen}},\ }\bibfield  {title} {\bibinfo {title} {Nonperturbative regularization of ($1+1$)-dimensional anomaly-free chiral fermions and bosons: On the equivalence of anomaly matching conditions and boundary gapping rules},\ }\href {https://doi.org/10.1103/PhysRevB.107.014311} {\bibfield  {journal} {\bibinfo  {journal} {Phys. Rev. B}\ }\textbf {\bibinfo {volume} {107}},\ \bibinfo {pages} {014311} (\bibinfo {year} {2023})}\BibitemShut {NoStop}%
\bibitem [{\citenamefont {Slagle}\ \emph {et~al.}(2015)\citenamefont {Slagle}, \citenamefont {You},\ and\ \citenamefont {Xu}}]{Slagle}%
  \BibitemOpen
  \bibfield  {author} {\bibinfo {author} {\bibfnamefont {K.}~\bibnamefont {Slagle}}, \bibinfo {author} {\bibfnamefont {Y.-Z.}\ \bibnamefont {You}},\ and\ \bibinfo {author} {\bibfnamefont {C.}~\bibnamefont {Xu}},\ }\bibfield  {title} {\bibinfo {title} {Exotic quantum phase transitions of strongly interacting topological insulators},\ }\href {https://doi.org/10.1103/PhysRevB.91.115121} {\bibfield  {journal} {\bibinfo  {journal} {Phys. Rev. B}\ }\textbf {\bibinfo {volume} {91}},\ \bibinfo {pages} {115121} (\bibinfo {year} {2015})}\BibitemShut {NoStop}%
\bibitem [{\citenamefont {Ayyar}\ and\ \citenamefont {Chandrasekharan}(2015)}]{shailesh}%
  \BibitemOpen
  \bibfield  {author} {\bibinfo {author} {\bibfnamefont {V.}~\bibnamefont {Ayyar}}\ and\ \bibinfo {author} {\bibfnamefont {S.}~\bibnamefont {Chandrasekharan}},\ }\bibfield  {title} {\bibinfo {title} {Massive fermions without fermion bilinear condensates},\ }\href {https://doi.org/10.1103/PhysRevD.91.065035} {\bibfield  {journal} {\bibinfo  {journal} {Phys. Rev. D}\ }\textbf {\bibinfo {volume} {91}},\ \bibinfo {pages} {065035} (\bibinfo {year} {2015})}\BibitemShut {NoStop}%
\bibitem [{\citenamefont {Ayyar}\ and\ \citenamefont {Chandrasekharan}(2016{\natexlab{a}})}]{shailesh2}%
  \BibitemOpen
  \bibfield  {author} {\bibinfo {author} {\bibfnamefont {V.}~\bibnamefont {Ayyar}}\ and\ \bibinfo {author} {\bibfnamefont {S.}~\bibnamefont {Chandrasekharan}},\ }\bibfield  {title} {\bibinfo {title} {Origin of fermion masses without spontaneous symmetry breaking},\ }\href {https://doi.org/10.1103/PhysRevD.93.081701} {\bibfield  {journal} {\bibinfo  {journal} {Phys. Rev. D}\ }\textbf {\bibinfo {volume} {93}},\ \bibinfo {pages} {081701} (\bibinfo {year} {2016}{\natexlab{a}})}\BibitemShut {NoStop}%
\bibitem [{\citenamefont {Ayyar}\ and\ \citenamefont {Chandrasekharan}(2016{\natexlab{b}})}]{shailesh3}%
  \BibitemOpen
  \bibfield  {author} {\bibinfo {author} {\bibfnamefont {V.}~\bibnamefont {Ayyar}}\ and\ \bibinfo {author} {\bibfnamefont {S.}~\bibnamefont {Chandrasekharan}},\ }\bibfield  {title} {\bibinfo {title} {Fermion masses through four-fermion condensates},\ }\bibfield  {journal} {\bibinfo  {journal} {Journal of High Energy Physics}\ }\textbf {\bibinfo {volume} {2016}},\ \href {https://doi.org/10.1007/jhep10(2016)058} {10.1007/jhep10(2016)058} (\bibinfo {year} {2016}{\natexlab{b}})\BibitemShut {NoStop}%
\bibitem [{\citenamefont {Catterall}(2016)}]{catterall_2016}%
  \BibitemOpen
  \bibfield  {author} {\bibinfo {author} {\bibfnamefont {S.}~\bibnamefont {Catterall}},\ }\bibfield  {title} {\bibinfo {title} {Fermion mass without symmetry breaking},\ }\bibfield  {journal} {\bibinfo  {journal} {Journal of High Energy Physics}\ }\textbf {\bibinfo {volume} {2016}},\ \href {https://doi.org/10.1007/jhep01(2016)121} {10.1007/jhep01(2016)121} (\bibinfo {year} {2016})\BibitemShut {NoStop}%
\bibitem [{\citenamefont {Catterall}\ and\ \citenamefont {Schaich}(2017)}]{catterall2}%
  \BibitemOpen
  \bibfield  {author} {\bibinfo {author} {\bibfnamefont {S.}~\bibnamefont {Catterall}}\ and\ \bibinfo {author} {\bibfnamefont {D.}~\bibnamefont {Schaich}},\ }\bibfield  {title} {\bibinfo {title} {Novel phases in strongly coupled four-fermion theories},\ }\href {https://doi.org/10.1103/PhysRevD.96.034506} {\bibfield  {journal} {\bibinfo  {journal} {Phys. Rev. D}\ }\textbf {\bibinfo {volume} {96}},\ \bibinfo {pages} {034506} (\bibinfo {year} {2017})}\BibitemShut {NoStop}%
\bibitem [{\citenamefont {Butt}\ \emph {et~al.}(2021)\citenamefont {Butt}, \citenamefont {Catterall},\ and\ \citenamefont {Toga}}]{catterall3}%
  \BibitemOpen
  \bibfield  {author} {\bibinfo {author} {\bibfnamefont {N.}~\bibnamefont {Butt}}, \bibinfo {author} {\bibfnamefont {S.}~\bibnamefont {Catterall}},\ and\ \bibinfo {author} {\bibfnamefont {G.~C.}\ \bibnamefont {Toga}},\ }\href {https://arxiv.org/abs/2111.01001} {\bibinfo {title} {Symmetric mass generation in lattice gauge theory}} (\bibinfo {year} {2021}),\ \Eprint {https://arxiv.org/abs/2111.01001} {arXiv:2111.01001 [hep-lat]} \BibitemShut {NoStop}%
\bibitem [{\citenamefont {Butt}\ \emph {et~al.}(2018)\citenamefont {Butt}, \citenamefont {Catterall},\ and\ \citenamefont {Schaich}}]{catterall4}%
  \BibitemOpen
  \bibfield  {author} {\bibinfo {author} {\bibfnamefont {N.}~\bibnamefont {Butt}}, \bibinfo {author} {\bibfnamefont {S.}~\bibnamefont {Catterall}},\ and\ \bibinfo {author} {\bibfnamefont {D.}~\bibnamefont {Schaich}},\ }\bibfield  {title} {\bibinfo {title} {$so\mathbf{(}4\mathbf{)}$ invariant higgs-yukawa model with reduced staggered fermions},\ }\href {https://doi.org/10.1103/PhysRevD.98.114514} {\bibfield  {journal} {\bibinfo  {journal} {Phys. Rev. D}\ }\textbf {\bibinfo {volume} {98}},\ \bibinfo {pages} {114514} (\bibinfo {year} {2018})}\BibitemShut {NoStop}%
\bibitem [{\citenamefont {He}\ \emph {et~al.}(2016)\citenamefont {He}, \citenamefont {Wu}, \citenamefont {You}, \citenamefont {Xu}, \citenamefont {Meng},\ and\ \citenamefont {Lu}}]{mengSMG1}%
  \BibitemOpen
  \bibfield  {author} {\bibinfo {author} {\bibfnamefont {Y.-Y.}\ \bibnamefont {He}}, \bibinfo {author} {\bibfnamefont {H.-Q.}\ \bibnamefont {Wu}}, \bibinfo {author} {\bibfnamefont {Y.-Z.}\ \bibnamefont {You}}, \bibinfo {author} {\bibfnamefont {C.}~\bibnamefont {Xu}}, \bibinfo {author} {\bibfnamefont {Z.~Y.}\ \bibnamefont {Meng}},\ and\ \bibinfo {author} {\bibfnamefont {Z.-Y.}\ \bibnamefont {Lu}},\ }\bibfield  {title} {\bibinfo {title} {Quantum critical point of dirac fermion mass generation without spontaneous symmetry breaking},\ }\href {https://doi.org/10.1103/PhysRevB.94.241111} {\bibfield  {journal} {\bibinfo  {journal} {Phys. Rev. B}\ }\textbf {\bibinfo {volume} {94}},\ \bibinfo {pages} {241111} (\bibinfo {year} {2016})}\BibitemShut {NoStop}%
\bibitem [{\citenamefont {Butt}\ \emph {et~al.}(2025)\citenamefont {Butt}, \citenamefont {Catterall},\ and\ \citenamefont {Hasenfratz}}]{annaSMG2}%
  \BibitemOpen
  \bibfield  {author} {\bibinfo {author} {\bibfnamefont {N.}~\bibnamefont {Butt}}, \bibinfo {author} {\bibfnamefont {S.}~\bibnamefont {Catterall}},\ and\ \bibinfo {author} {\bibfnamefont {A.}~\bibnamefont {Hasenfratz}},\ }\bibfield  {title} {\bibinfo {title} {Symmetric mass generation with four su(2) doublet fermions},\ }\href {https://doi.org/10.1103/PhysRevLett.134.031602} {\bibfield  {journal} {\bibinfo  {journal} {Phys. Rev. Lett.}\ }\textbf {\bibinfo {volume} {134}},\ \bibinfo {pages} {031602} (\bibinfo {year} {2025})}\BibitemShut {NoStop}%
\bibitem [{\citenamefont {You}\ \emph {et~al.}(2018{\natexlab{a}})\citenamefont {You}, \citenamefont {He}, \citenamefont {Xu},\ and\ \citenamefont {Vishwanath}}]{SMGDQCP}%
  \BibitemOpen
  \bibfield  {author} {\bibinfo {author} {\bibfnamefont {Y.-Z.}\ \bibnamefont {You}}, \bibinfo {author} {\bibfnamefont {Y.-C.}\ \bibnamefont {He}}, \bibinfo {author} {\bibfnamefont {C.}~\bibnamefont {Xu}},\ and\ \bibinfo {author} {\bibfnamefont {A.}~\bibnamefont {Vishwanath}},\ }\bibfield  {title} {\bibinfo {title} {Symmetric fermion mass generation as deconfined quantum criticality},\ }\href {https://doi.org/10.1103/PhysRevX.8.011026} {\bibfield  {journal} {\bibinfo  {journal} {Phys. Rev. X}\ }\textbf {\bibinfo {volume} {8}},\ \bibinfo {pages} {011026} (\bibinfo {year} {2018}{\natexlab{a}})}\BibitemShut {NoStop}%
\bibitem [{\citenamefont {You}\ \emph {et~al.}(2018{\natexlab{b}})\citenamefont {You}, \citenamefont {He}, \citenamefont {Vishwanath},\ and\ \citenamefont {Xu}}]{SMGDQCP2}%
  \BibitemOpen
  \bibfield  {author} {\bibinfo {author} {\bibfnamefont {Y.-Z.}\ \bibnamefont {You}}, \bibinfo {author} {\bibfnamefont {Y.-C.}\ \bibnamefont {He}}, \bibinfo {author} {\bibfnamefont {A.}~\bibnamefont {Vishwanath}},\ and\ \bibinfo {author} {\bibfnamefont {C.}~\bibnamefont {Xu}},\ }\bibfield  {title} {\bibinfo {title} {From bosonic topological transition to symmetric fermion mass generation},\ }\href {https://doi.org/10.1103/PhysRevB.97.125112} {\bibfield  {journal} {\bibinfo  {journal} {Phys. Rev. B}\ }\textbf {\bibinfo {volume} {97}},\ \bibinfo {pages} {125112} (\bibinfo {year} {2018}{\natexlab{b}})}\BibitemShut {NoStop}%
\bibitem [{\citenamefont {Razamat}\ and\ \citenamefont {Tong}(2021)}]{Tong_2021}%
  \BibitemOpen
  \bibfield  {author} {\bibinfo {author} {\bibfnamefont {S.~S.}\ \bibnamefont {Razamat}}\ and\ \bibinfo {author} {\bibfnamefont {D.}~\bibnamefont {Tong}},\ }\bibfield  {title} {\bibinfo {title} {Gapped chiral fermions},\ }\bibfield  {journal} {\bibinfo  {journal} {Physical Review X}\ }\textbf {\bibinfo {volume} {11}},\ \href {https://doi.org/10.1103/physrevx.11.011063} {10.1103/physrevx.11.011063} (\bibinfo {year} {2021})\BibitemShut {NoStop}%
\bibitem [{\citenamefont {Tong}(2022)}]{Tong_2022}%
  \BibitemOpen
  \bibfield  {author} {\bibinfo {author} {\bibfnamefont {D.}~\bibnamefont {Tong}},\ }\bibfield  {title} {\bibinfo {title} {Comments on symmetric mass generation in 2d and 4d},\ }\bibfield  {journal} {\bibinfo  {journal} {Journal of High Energy Physics}\ }\textbf {\bibinfo {volume} {2022}},\ \href {https://doi.org/10.1007/jhep07(2022)001} {10.1007/jhep07(2022)001} (\bibinfo {year} {2022})\BibitemShut {NoStop}%
\bibitem [{\citenamefont {Zeng}\ \emph {et~al.}(2022)\citenamefont {Zeng}, \citenamefont {Zhu}, \citenamefont {Wang},\ and\ \citenamefont {You}}]{34502}%
  \BibitemOpen
  \bibfield  {author} {\bibinfo {author} {\bibfnamefont {M.}~\bibnamefont {Zeng}}, \bibinfo {author} {\bibfnamefont {Z.}~\bibnamefont {Zhu}}, \bibinfo {author} {\bibfnamefont {J.}~\bibnamefont {Wang}},\ and\ \bibinfo {author} {\bibfnamefont {Y.-Z.}\ \bibnamefont {You}},\ }\bibfield  {title} {\bibinfo {title} {Symmetric mass generation in the $1+1$ dimensional chiral fermion 3-4-5-0 model},\ }\href {https://doi.org/10.1103/PhysRevLett.128.185301} {\bibfield  {journal} {\bibinfo  {journal} {Phys. Rev. Lett.}\ }\textbf {\bibinfo {volume} {128}},\ \bibinfo {pages} {185301} (\bibinfo {year} {2022})}\BibitemShut {NoStop}%
\bibitem [{\citenamefont {Liu}\ \emph {et~al.}(2024)\citenamefont {Liu}, \citenamefont {Da~Liao}, \citenamefont {Pan}, \citenamefont {Song}, \citenamefont {Zhao}, \citenamefont {Jiang}, \citenamefont {Jian}, \citenamefont {You}, \citenamefont {Assaad}, \citenamefont {Meng},\ and\ \citenamefont {Xu}}]{disorderSMG}%
  \BibitemOpen
  \bibfield  {author} {\bibinfo {author} {\bibfnamefont {Z.~H.}\ \bibnamefont {Liu}}, \bibinfo {author} {\bibfnamefont {Y.}~\bibnamefont {Da~Liao}}, \bibinfo {author} {\bibfnamefont {G.}~\bibnamefont {Pan}}, \bibinfo {author} {\bibfnamefont {M.}~\bibnamefont {Song}}, \bibinfo {author} {\bibfnamefont {J.}~\bibnamefont {Zhao}}, \bibinfo {author} {\bibfnamefont {W.}~\bibnamefont {Jiang}}, \bibinfo {author} {\bibfnamefont {C.-M.}\ \bibnamefont {Jian}}, \bibinfo {author} {\bibfnamefont {Y.-Z.}\ \bibnamefont {You}}, \bibinfo {author} {\bibfnamefont {F.~F.}\ \bibnamefont {Assaad}}, \bibinfo {author} {\bibfnamefont {Z.~Y.}\ \bibnamefont {Meng}},\ and\ \bibinfo {author} {\bibfnamefont {C.}~\bibnamefont {Xu}},\ }\bibfield  {title} {\bibinfo {title} {Disorder operator and r\'enyi entanglement entropy of symmetric mass generation},\ }\href {https://doi.org/10.1103/PhysRevLett.132.156503} {\bibfield  {journal} {\bibinfo  {journal} {Phys. Rev. Lett.}\ }\textbf {\bibinfo {volume} {132}},\ \bibinfo {pages} {156503} (\bibinfo
  {year} {2024})}\BibitemShut {NoStop}%
\bibitem [{\citenamefont {Wang}\ and\ \citenamefont {You}(2022)}]{SMGreview}%
  \BibitemOpen
  \bibfield  {author} {\bibinfo {author} {\bibfnamefont {J.}~\bibnamefont {Wang}}\ and\ \bibinfo {author} {\bibfnamefont {Y.-Z.}\ \bibnamefont {You}},\ }\bibfield  {title} {\bibinfo {title} {Symmetric mass generation},\ }\href {https://doi.org/10.3390/sym14071475} {\bibfield  {journal} {\bibinfo  {journal} {Symmetry}\ }\textbf {\bibinfo {volume} {14}},\ \bibinfo {pages} {1475} (\bibinfo {year} {2022})}\BibitemShut {NoStop}%
\bibitem [{\citenamefont {Hasenfratz}\ and\ \citenamefont {Witzel}(2024)}]{annaSMG1}%
  \BibitemOpen
  \bibfield  {author} {\bibinfo {author} {\bibfnamefont {A.}~\bibnamefont {Hasenfratz}}\ and\ \bibinfo {author} {\bibfnamefont {O.}~\bibnamefont {Witzel}},\ }\href {https://arxiv.org/abs/2412.10322} {\bibinfo {title} {Investigating su(3) with nf=8 fundamental fermions at strong renormalized coupling}} (\bibinfo {year} {2024}),\ \Eprint {https://arxiv.org/abs/2412.10322} {arXiv:2412.10322 [hep-lat]} \BibitemShut {NoStop}%
\bibitem [{\citenamefont {Maiti}\ \emph {et~al.}(2025)\citenamefont {Maiti}, \citenamefont {Banerjee}, \citenamefont {Chandrasekharan},\ and\ \citenamefont {Marinkovic}}]{marina1}%
  \BibitemOpen
  \bibfield  {author} {\bibinfo {author} {\bibfnamefont {S.}~\bibnamefont {Maiti}}, \bibinfo {author} {\bibfnamefont {D.}~\bibnamefont {Banerjee}}, \bibinfo {author} {\bibfnamefont {S.}~\bibnamefont {Chandrasekharan}},\ and\ \bibinfo {author} {\bibfnamefont {M.~K.}\ \bibnamefont {Marinkovic}},\ }\href {https://arxiv.org/abs/2512.24836} {\bibinfo {title} {Symmetric mass generation as a multicritical point with enhanced symmetry}} (\bibinfo {year} {2025}),\ \Eprint {https://arxiv.org/abs/2512.24836} {arXiv:2512.24836 [hep-lat]} \BibitemShut {NoStop}%
\bibitem [{\citenamefont {Maiti}\ \emph {et~al.}(2026)\citenamefont {Maiti}, \citenamefont {Banerjee}, \citenamefont {Chandrasekharan},\ and\ \citenamefont {Marinkovic}}]{marina2}%
  \BibitemOpen
  \bibfield  {author} {\bibinfo {author} {\bibfnamefont {S.}~\bibnamefont {Maiti}}, \bibinfo {author} {\bibfnamefont {D.}~\bibnamefont {Banerjee}}, \bibinfo {author} {\bibfnamefont {S.}~\bibnamefont {Chandrasekharan}},\ and\ \bibinfo {author} {\bibfnamefont {M.~K.}\ \bibnamefont {Marinkovic}},\ }\href {https://arxiv.org/abs/2602.18360} {\bibinfo {title} {Phase diagram of a lattice fermion model with symmetric mass generation}} (\bibinfo {year} {2026}),\ \Eprint {https://arxiv.org/abs/2602.18360} {arXiv:2602.18360 [hep-lat]} \BibitemShut {NoStop}%
\bibitem [{\citenamefont {Wen}(2013)}]{wenSM}%
  \BibitemOpen
  \bibfield  {author} {\bibinfo {author} {\bibfnamefont {X.-G.}\ \bibnamefont {Wen}},\ }\href@noop {} {\bibfield  {journal} {\bibinfo  {journal} {Chin. Phys. Lett.}\ }\textbf {\bibinfo {volume} {30}},\ \bibinfo {pages} {111101} (\bibinfo {year} {2013})}\BibitemShut {NoStop}%
\bibitem [{\citenamefont {You}\ \emph {et~al.}(2014)\citenamefont {You}, \citenamefont {BenTov},\ and\ \citenamefont {Xu}}]{xu16}%
  \BibitemOpen
  \bibfield  {author} {\bibinfo {author} {\bibfnamefont {Y.-Z.}\ \bibnamefont {You}}, \bibinfo {author} {\bibfnamefont {Y.}~\bibnamefont {BenTov}},\ and\ \bibinfo {author} {\bibfnamefont {C.}~\bibnamefont {Xu}},\ }\href {https://arxiv.org/abs/1402.4151} {\bibinfo {title} {Interacting topological superconductors and possible origin of $16n$ chiral fermions in the standard model}} (\bibinfo {year} {2014}),\ \Eprint {https://arxiv.org/abs/1402.4151} {arXiv:1402.4151 [cond-mat.str-el]} \BibitemShut {NoStop}%
\bibitem [{\citenamefont {You}\ and\ \citenamefont {Xu}(2015)}]{xuGUT}%
  \BibitemOpen
  \bibfield  {author} {\bibinfo {author} {\bibfnamefont {Y.-Z.}\ \bibnamefont {You}}\ and\ \bibinfo {author} {\bibfnamefont {C.}~\bibnamefont {Xu}},\ }\href@noop {} {\bibfield  {journal} {\bibinfo  {journal} {Phys. Rev. B}\ }\textbf {\bibinfo {volume} {91}},\ \bibinfo {pages} {125147} (\bibinfo {year} {2015})}\BibitemShut {NoStop}%
\bibitem [{\citenamefont {Wang}\ and\ \citenamefont {Wen}(2020)}]{Wang_2020}%
  \BibitemOpen
  \bibfield  {author} {\bibinfo {author} {\bibfnamefont {J.}~\bibnamefont {Wang}}\ and\ \bibinfo {author} {\bibfnamefont {X.-G.}\ \bibnamefont {Wen}},\ }\bibfield  {title} {\bibinfo {title} {Nonperturbative definition of the standard models},\ }\bibfield  {journal} {\bibinfo  {journal} {Physical Review Research}\ }\textbf {\bibinfo {volume} {2}},\ \href {https://doi.org/10.1103/physrevresearch.2.023356} {10.1103/physrevresearch.2.023356} (\bibinfo {year} {2020})\BibitemShut {NoStop}%
\bibitem [{\citenamefont {Tachikawa}\ and\ \citenamefont {Yonekura}(2019)}]{spinZ41}%
  \BibitemOpen
  \bibfield  {author} {\bibinfo {author} {\bibfnamefont {Y.}~\bibnamefont {Tachikawa}}\ and\ \bibinfo {author} {\bibfnamefont {K.}~\bibnamefont {Yonekura}},\ }\bibfield  {title} {\bibinfo {title} {Why are fractional charges of orientifolds compatible with dirac quantization?},\ }\bibfield  {journal} {\bibinfo  {journal} {SciPost Physics}\ }\textbf {\bibinfo {volume} {7}},\ \href {https://doi.org/10.21468/scipostphys.7.5.058} {10.21468/scipostphys.7.5.058} (\bibinfo {year} {2019})\BibitemShut {NoStop}%
\bibitem [{\citenamefont {Garcia-Etxebarria}\ and\ \citenamefont {Montero}(2019)}]{spinZ42}%
  \BibitemOpen
  \bibfield  {author} {\bibinfo {author} {\bibfnamefont {I.}~\bibnamefont {Garcia-Etxebarria}}\ and\ \bibinfo {author} {\bibfnamefont {M.}~\bibnamefont {Montero}},\ }\bibfield  {title} {\bibinfo {title} {Dai-freed anomalies in particle physics},\ }\bibfield  {journal} {\bibinfo  {journal} {Journal of High Energy Physics}\ }\textbf {\bibinfo {volume} {2019}},\ \href {https://doi.org/10.1007/jhep08(2019)003} {10.1007/jhep08(2019)003} (\bibinfo {year} {2019})\BibitemShut {NoStop}%
\bibitem [{\citenamefont {Catterall}(2023)}]{spinZ43}%
  \BibitemOpen
  \bibfield  {author} {\bibinfo {author} {\bibfnamefont {S.}~\bibnamefont {Catterall}},\ }\bibfield  {title} {\bibinfo {title} {'t hooft anomalies for staggered fermions},\ }\bibfield  {journal} {\bibinfo  {journal} {Physical Review D}\ }\textbf {\bibinfo {volume} {107}},\ \href {https://doi.org/10.1103/physrevd.107.014501} {10.1103/physrevd.107.014501} (\bibinfo {year} {2023})\BibitemShut {NoStop}%
\bibitem [{Note1()}]{Note1}%
  \BibitemOpen
  \bibinfo {note} {An onsite symmetry means that the symmetry operator is the product of operators acting on local tensor-product Hilbert spaces; it also implies that the symmetry can be promoted to a gauge symmetry. A non-onsite symmetry would generally involve lattice transformations, such as translation.}\BibitemShut {Stop}%
\bibitem [{\citenamefont {Nielsen}\ and\ \citenamefont {Ninomiya}(1981{\natexlab{a}})}]{nn1}%
  \BibitemOpen
  \bibfield  {author} {\bibinfo {author} {\bibfnamefont {H.}~\bibnamefont {Nielsen}}\ and\ \bibinfo {author} {\bibfnamefont {M.}~\bibnamefont {Ninomiya}},\ }\bibfield  {title} {\bibinfo {title} {Absence of neutrinos on a lattice: (i). proof by homotopy theory},\ }\href {https://doi.org/https://doi.org/10.1016/0550-3213(81)90361-8} {\bibfield  {journal} {\bibinfo  {journal} {Nuclear Physics B}\ }\textbf {\bibinfo {volume} {185}},\ \bibinfo {pages} {20} (\bibinfo {year} {1981}{\natexlab{a}})}\BibitemShut {NoStop}%
\bibitem [{\citenamefont {Nielsen}\ and\ \citenamefont {Ninomiya}(1981{\natexlab{b}})}]{nn2}%
  \BibitemOpen
  \bibfield  {author} {\bibinfo {author} {\bibfnamefont {H.}~\bibnamefont {Nielsen}}\ and\ \bibinfo {author} {\bibfnamefont {M.}~\bibnamefont {Ninomiya}},\ }\bibfield  {title} {\bibinfo {title} {Absence of neutrinos on a lattice: (ii). intuitive topological proof},\ }\href {https://doi.org/https://doi.org/10.1016/0550-3213(81)90524-1} {\bibfield  {journal} {\bibinfo  {journal} {Nuclear Physics B}\ }\textbf {\bibinfo {volume} {193}},\ \bibinfo {pages} {173} (\bibinfo {year} {1981}{\natexlab{b}})}\BibitemShut {NoStop}%
\bibitem [{\citenamefont {Shamir}(1993)}]{shamir1}%
  \BibitemOpen
  \bibfield  {author} {\bibinfo {author} {\bibfnamefont {Y.}~\bibnamefont {Shamir}},\ }\bibfield  {title} {\bibinfo {title} {Constraints on the existence of chiral fermions in interacting lattice theories},\ }\href@noop {} {\bibfield  {journal} {\bibinfo  {journal} {Phys. Rev. Lett.}\ }\textbf {\bibinfo {volume} {71}},\ \bibinfo {pages} {2691} (\bibinfo {year} {1993})}\BibitemShut {NoStop}%
\bibitem [{\citenamefont {Golterman}\ and\ \citenamefont {Shamir}(2025)}]{shamir2}%
  \BibitemOpen
  \bibfield  {author} {\bibinfo {author} {\bibfnamefont {M.}~\bibnamefont {Golterman}}\ and\ \bibinfo {author} {\bibfnamefont {Y.}~\bibnamefont {Shamir}},\ }\href {https://arxiv.org/abs/2505.20436} {\bibinfo {title} {Constraints on the symmetric mass generation paradigm for lattice chiral gauge theories}} (\bibinfo {year} {2025}),\ \Eprint {https://arxiv.org/abs/2505.20436} {arXiv:2505.20436 [hep-lat]} \BibitemShut {NoStop}%
\bibitem [{\citenamefont {Fidkowski}\ and\ \citenamefont {Xu}(2023)}]{fidkowskixu}%
  \BibitemOpen
  \bibfield  {author} {\bibinfo {author} {\bibfnamefont {L.}~\bibnamefont {Fidkowski}}\ and\ \bibinfo {author} {\bibfnamefont {C.}~\bibnamefont {Xu}},\ }\bibfield  {title} {\bibinfo {title} {A no-go result for implementing chiral symmetries by locality-preserving unitaries in a three-dimensional hamiltonian lattice model of fermions},\ }\href {https://doi.org/10.1103/PhysRevLett.131.196601} {\bibfield  {journal} {\bibinfo  {journal} {Phys. Rev. Lett.}\ }\textbf {\bibinfo {volume} {131}},\ \bibinfo {pages} {196601} (\bibinfo {year} {2023})}\BibitemShut {NoStop}%
\bibitem [{\citenamefont {Tanizaki}(2018)}]{newanomaly}%
  \BibitemOpen
  \bibfield  {author} {\bibinfo {author} {\bibfnamefont {Y.}~\bibnamefont {Tanizaki}},\ }\bibfield  {title} {\bibinfo {title} {Anomaly constraint on massless qcd and the role of skyrmions in chiral symmetry breaking},\ }\bibfield  {journal} {\bibinfo  {journal} {Journal of High Energy Physics}\ }\textbf {\bibinfo {volume} {2018}},\ \href {https://doi.org/10.1007/jhep08(2018)171} {10.1007/jhep08(2018)171} (\bibinfo {year} {2018})\BibitemShut {NoStop}%
\bibitem [{\citenamefont {Córdova}\ \emph {et~al.}(2020)\citenamefont {Córdova}, \citenamefont {Freed}, \citenamefont {Lam},\ and\ \citenamefont {Seiberg}}]{newanomaly1}%
  \BibitemOpen
  \bibfield  {author} {\bibinfo {author} {\bibfnamefont {C.}~\bibnamefont {Córdova}}, \bibinfo {author} {\bibfnamefont {D.~S.}\ \bibnamefont {Freed}}, \bibinfo {author} {\bibfnamefont {H.~T.}\ \bibnamefont {Lam}},\ and\ \bibinfo {author} {\bibfnamefont {N.}~\bibnamefont {Seiberg}},\ }\bibfield  {title} {\bibinfo {title} {{Anomalies in the space of coupling constants and their dynamical applications II}},\ }\href {https://doi.org/10.21468/SciPostPhys.8.1.002} {\bibfield  {journal} {\bibinfo  {journal} {SciPost Phys.}\ }\textbf {\bibinfo {volume} {8}},\ \bibinfo {pages} {002} (\bibinfo {year} {2020})}\BibitemShut {NoStop}%
\bibitem [{\citenamefont {Brennan}\ and\ \citenamefont {Cordova}(2021)}]{newanomaly2}%
  \BibitemOpen
  \bibfield  {author} {\bibinfo {author} {\bibfnamefont {T.~D.}\ \bibnamefont {Brennan}}\ and\ \bibinfo {author} {\bibfnamefont {C.}~\bibnamefont {Cordova}},\ }\href {https://arxiv.org/abs/2011.09600} {\bibinfo {title} {Axions, higher-groups, and emergent symmetry}} (\bibinfo {year} {2021}),\ \Eprint {https://arxiv.org/abs/2011.09600} {arXiv:2011.09600 [hep-th]} \BibitemShut {NoStop}%
\bibitem [{\citenamefont {Gaiotto}\ \emph {et~al.}(2018)\citenamefont {Gaiotto}, \citenamefont {Komargodski},\ and\ \citenamefont {Seiberg}}]{zohar2017}%
  \BibitemOpen
  \bibfield  {author} {\bibinfo {author} {\bibfnamefont {D.}~\bibnamefont {Gaiotto}}, \bibinfo {author} {\bibfnamefont {Z.}~\bibnamefont {Komargodski}},\ and\ \bibinfo {author} {\bibfnamefont {N.}~\bibnamefont {Seiberg}},\ }\bibfield  {title} {\bibinfo {title} {Time-reversal breaking in qcd4, walls, and dualities in 2 + 1 dimensions},\ }\bibfield  {journal} {\bibinfo  {journal} {Journal of High Energy Physics}\ }\textbf {\bibinfo {volume} {2018}},\ \href {https://doi.org/10.1007/jhep01(2018)110} {10.1007/jhep01(2018)110} (\bibinfo {year} {2018})\BibitemShut {NoStop}%
\bibitem [{\citenamefont {Lam}\ \emph {et~al.}(2026)\citenamefont {Lam}, \citenamefont {Sehayek},\ and\ \citenamefont {Xu}}]{future}%
  \BibitemOpen
  \bibfield  {author} {\bibinfo {author} {\bibfnamefont {H.~T.}\ \bibnamefont {Lam}}, \bibinfo {author} {\bibfnamefont {D.}~\bibnamefont {Sehayek}},\ and\ \bibinfo {author} {\bibfnamefont {C.}~\bibnamefont {Xu}},\ }\bibfield  {title} {\bibinfo {title} {Anomaly analysis for symmetric mass generation},\ }\href@noop {} {\  (\bibinfo {year} {2026})},\ \bibinfo {note} {to appear}\BibitemShut {NoStop}%
\bibitem [{\citenamefont {Lee}\ and\ \citenamefont {Sharpe}(1999)}]{leesharpe}%
  \BibitemOpen
  \bibfield  {author} {\bibinfo {author} {\bibfnamefont {W.}~\bibnamefont {Lee}}\ and\ \bibinfo {author} {\bibfnamefont {S.~R.}\ \bibnamefont {Sharpe}},\ }\bibfield  {title} {\bibinfo {title} {Partial flavor symmetry restoration for chiral staggered fermions},\ }\href {https://doi.org/10.1103/PhysRevD.60.114503} {\bibfield  {journal} {\bibinfo  {journal} {Phys. Rev. D}\ }\textbf {\bibinfo {volume} {60}},\ \bibinfo {pages} {114503} (\bibinfo {year} {1999})}\BibitemShut {NoStop}%
\bibitem [{\citenamefont {Weingarten}(1983)}]{weingarten}%
  \BibitemOpen
  \bibfield  {author} {\bibinfo {author} {\bibfnamefont {D.}~\bibnamefont {Weingarten}},\ }\bibfield  {title} {\bibinfo {title} {Mass inequalities for quantum chromodynamics},\ }\href {https://doi.org/10.1103/PhysRevLett.51.1830} {\bibfield  {journal} {\bibinfo  {journal} {Phys. Rev. Lett.}\ }\textbf {\bibinfo {volume} {51}},\ \bibinfo {pages} {1830} (\bibinfo {year} {1983})}\BibitemShut {NoStop}%
\bibitem [{Note2()}]{Note2}%
  \BibitemOpen
  \bibinfo {note} {In fact, if ${\mathchoice {{\setbox \z@ \hbox {$\mathsurround \z@ \displaystyle D$}\setbox \tw@ \hbox {$\mathsurround \z@ \displaystyle /$}\dimen 4\wd \z@ \dimen@ \ht \tw@ \advance \dimen@ -\dp \tw@ \advance \dimen@ -\ht \z@ \advance \dimen@ \dp \z@ \divide \dimen@ \tw@ \advance \dimen@ -0\ht \tw@ \advance \dimen@ -0\dp \tw@ \dimen@ii .08\wd \z@ \raise -\dimen@ \hbox to\dimen 4{\hss \kern \dimen@ii \box \tw@ \kern -\dimen@ii \hss }\protect \llap {\hbox to\dimen 4{\hss \box \z@ \hss }}}}{{\setbox \z@ \hbox {$\mathsurround \z@ \textstyle D$}\setbox \tw@ \hbox {$\mathsurround \z@ \textstyle /$}\dimen 4\wd \z@ \dimen@ \ht \tw@ \advance \dimen@ -\dp \tw@ \advance \dimen@ -\ht \z@ \advance \dimen@ \dp \z@ \divide \dimen@ \tw@ \advance \dimen@ -0\ht \tw@ \advance \dimen@ -0\dp \tw@ \dimen@ii .08\wd \z@ \raise -\dimen@ \hbox to\dimen 4{\hss \kern \dimen@ii \box \tw@ \kern -\dimen@ii \hss }\protect \llap {\hbox to\dimen 4{\hss \box \z@ \hss }}}}{{\setbox \z@ \hbox {$\mathsurround \z@ \scriptstyle
  D$}\setbox \tw@ \hbox {$\mathsurround \z@ \scriptstyle /$}\dimen 4\wd \z@ \dimen@ \ht \tw@ \advance \dimen@ -\dp \tw@ \advance \dimen@ -\ht \z@ \advance \dimen@ \dp \z@ \divide \dimen@ \tw@ \advance \dimen@ -0\ht \tw@ \advance \dimen@ -0\dp \tw@ \dimen@ii .08\wd \z@ \raise -\dimen@ \hbox to\dimen 4{\hss \kern \dimen@ii \box \tw@ \kern -\dimen@ii \hss }\protect \llap {\hbox to\dimen 4{\hss \box \z@ \hss }}}}{{\setbox \z@ \hbox {$\mathsurround \z@ \scriptscriptstyle D$}\setbox \tw@ \hbox {$\mathsurround \z@ \scriptscriptstyle /$}\dimen 4\wd \z@ \dimen@ \ht \tw@ \advance \dimen@ -\dp \tw@ \advance \dimen@ -\ht \z@ \advance \dimen@ \dp \z@ \divide \dimen@ \tw@ \advance \dimen@ -0\ht \tw@ \advance \dimen@ -0\dp \tw@ \dimen@ii .08\wd \z@ \raise -\dimen@ \hbox to\dimen 4{\hss \kern \dimen@ii \box \tw@ \kern -\dimen@ii \hss }\protect \llap {\hbox to\dimen 4{\hss \box \z@ \hss }}}}}_{A,0}$ has no zero mode, ${\protect \rm det} {\mathchoice {{\setbox \z@ \hbox {$\mathsurround \z@ \displaystyle D$}\setbox \tw@ \hbox
  {$\mathsurround \z@ \displaystyle /$}\dimen 4\wd \z@ \dimen@ \ht \tw@ \advance \dimen@ -\dp \tw@ \advance \dimen@ -\ht \z@ \advance \dimen@ \dp \z@ \divide \dimen@ \tw@ \advance \dimen@ -0\ht \tw@ \advance \dimen@ -0\dp \tw@ \dimen@ii .08\wd \z@ \raise -\dimen@ \hbox to\dimen 4{\hss \kern \dimen@ii \box \tw@ \kern -\dimen@ii \hss }\protect \llap {\hbox to\dimen 4{\hss \box \z@ \hss }}}}{{\setbox \z@ \hbox {$\mathsurround \z@ \textstyle D$}\setbox \tw@ \hbox {$\mathsurround \z@ \textstyle /$}\dimen 4\wd \z@ \dimen@ \ht \tw@ \advance \dimen@ -\dp \tw@ \advance \dimen@ -\ht \z@ \advance \dimen@ \dp \z@ \divide \dimen@ \tw@ \advance \dimen@ -0\ht \tw@ \advance \dimen@ -0\dp \tw@ \dimen@ii .08\wd \z@ \raise -\dimen@ \hbox to\dimen 4{\hss \kern \dimen@ii \box \tw@ \kern -\dimen@ii \hss }\protect \llap {\hbox to\dimen 4{\hss \box \z@ \hss }}}}{{\setbox \z@ \hbox {$\mathsurround \z@ \scriptstyle D$}\setbox \tw@ \hbox {$\mathsurround \z@ \scriptstyle /$}\dimen 4\wd \z@ \dimen@ \ht \tw@ \advance \dimen@ -\dp \tw@
  \advance \dimen@ -\ht \z@ \advance \dimen@ \dp \z@ \divide \dimen@ \tw@ \advance \dimen@ -0\ht \tw@ \advance \dimen@ -0\dp \tw@ \dimen@ii .08\wd \z@ \raise -\dimen@ \hbox to\dimen 4{\hss \kern \dimen@ii \box \tw@ \kern -\dimen@ii \hss }\protect \llap {\hbox to\dimen 4{\hss \box \z@ \hss }}}}{{\setbox \z@ \hbox {$\mathsurround \z@ \scriptscriptstyle D$}\setbox \tw@ \hbox {$\mathsurround \z@ \scriptscriptstyle /$}\dimen 4\wd \z@ \dimen@ \ht \tw@ \advance \dimen@ -\dp \tw@ \advance \dimen@ -\ht \z@ \advance \dimen@ \dp \z@ \divide \dimen@ \tw@ \advance \dimen@ -0\ht \tw@ \advance \dimen@ -0\dp \tw@ \dimen@ii .08\wd \z@ \raise -\dimen@ \hbox to\dimen 4{\hss \kern \dimen@ii \box \tw@ \kern -\dimen@ii \hss }\protect \llap {\hbox to\dimen 4{\hss \box \z@ \hss }}}}}_A$ is already positive. If we take into account of zero modes of ${\mathchoice {{\setbox \z@ \hbox {$\mathsurround \z@ \displaystyle D$}\setbox \tw@ \hbox {$\mathsurround \z@ \displaystyle /$}\dimen 4\wd \z@ \dimen@ \ht \tw@ \advance \dimen@ -\dp \tw@
  \advance \dimen@ -\ht \z@ \advance \dimen@ \dp \z@ \divide \dimen@ \tw@ \advance \dimen@ -0\ht \tw@ \advance \dimen@ -0\dp \tw@ \dimen@ii .08\wd \z@ \raise -\dimen@ \hbox to\dimen 4{\hss \kern \dimen@ii \box \tw@ \kern -\dimen@ii \hss }\protect \llap {\hbox to\dimen 4{\hss \box \z@ \hss }}}}{{\setbox \z@ \hbox {$\mathsurround \z@ \textstyle D$}\setbox \tw@ \hbox {$\mathsurround \z@ \textstyle /$}\dimen 4\wd \z@ \dimen@ \ht \tw@ \advance \dimen@ -\dp \tw@ \advance \dimen@ -\ht \z@ \advance \dimen@ \dp \z@ \divide \dimen@ \tw@ \advance \dimen@ -0\ht \tw@ \advance \dimen@ -0\dp \tw@ \dimen@ii .08\wd \z@ \raise -\dimen@ \hbox to\dimen 4{\hss \kern \dimen@ii \box \tw@ \kern -\dimen@ii \hss }\protect \llap {\hbox to\dimen 4{\hss \box \z@ \hss }}}}{{\setbox \z@ \hbox {$\mathsurround \z@ \scriptstyle D$}\setbox \tw@ \hbox {$\mathsurround \z@ \scriptstyle /$}\dimen 4\wd \z@ \dimen@ \ht \tw@ \advance \dimen@ -\dp \tw@ \advance \dimen@ -\ht \z@ \advance \dimen@ \dp \z@ \divide \dimen@ \tw@ \advance \dimen@ -0\ht \tw@
  \advance \dimen@ -0\dp \tw@ \dimen@ii .08\wd \z@ \raise -\dimen@ \hbox to\dimen 4{\hss \kern \dimen@ii \box \tw@ \kern -\dimen@ii \hss }\protect \llap {\hbox to\dimen 4{\hss \box \z@ \hss }}}}{{\setbox \z@ \hbox {$\mathsurround \z@ \scriptscriptstyle D$}\setbox \tw@ \hbox {$\mathsurround \z@ \scriptscriptstyle /$}\dimen 4\wd \z@ \dimen@ \ht \tw@ \advance \dimen@ -\dp \tw@ \advance \dimen@ -\ht \z@ \advance \dimen@ \dp \z@ \divide \dimen@ \tw@ \advance \dimen@ -0\ht \tw@ \advance \dimen@ -0\dp \tw@ \dimen@ii .08\wd \z@ \raise -\dimen@ \hbox to\dimen 4{\hss \kern \dimen@ii \box \tw@ \kern -\dimen@ii \hss }\protect \llap {\hbox to\dimen 4{\hss \box \z@ \hss }}}}}_{A,0}$, $({\protect \rm det} {\mathchoice {{\setbox \z@ \hbox {$\mathsurround \z@ \displaystyle D$}\setbox \tw@ \hbox {$\mathsurround \z@ \displaystyle /$}\dimen 4\wd \z@ \dimen@ \ht \tw@ \advance \dimen@ -\dp \tw@ \advance \dimen@ -\ht \z@ \advance \dimen@ \dp \z@ \divide \dimen@ \tw@ \advance \dimen@ -0\ht \tw@ \advance \dimen@ -0\dp \tw@ \dimen@ii
  .08\wd \z@ \raise -\dimen@ \hbox to\dimen 4{\hss \kern \dimen@ii \box \tw@ \kern -\dimen@ii \hss }\protect \llap {\hbox to\dimen 4{\hss \box \z@ \hss }}}}{{\setbox \z@ \hbox {$\mathsurround \z@ \textstyle D$}\setbox \tw@ \hbox {$\mathsurround \z@ \textstyle /$}\dimen 4\wd \z@ \dimen@ \ht \tw@ \advance \dimen@ -\dp \tw@ \advance \dimen@ -\ht \z@ \advance \dimen@ \dp \z@ \divide \dimen@ \tw@ \advance \dimen@ -0\ht \tw@ \advance \dimen@ -0\dp \tw@ \dimen@ii .08\wd \z@ \raise -\dimen@ \hbox to\dimen 4{\hss \kern \dimen@ii \box \tw@ \kern -\dimen@ii \hss }\protect \llap {\hbox to\dimen 4{\hss \box \z@ \hss }}}}{{\setbox \z@ \hbox {$\mathsurround \z@ \scriptstyle D$}\setbox \tw@ \hbox {$\mathsurround \z@ \scriptstyle /$}\dimen 4\wd \z@ \dimen@ \ht \tw@ \advance \dimen@ -\dp \tw@ \advance \dimen@ -\ht \z@ \advance \dimen@ \dp \z@ \divide \dimen@ \tw@ \advance \dimen@ -0\ht \tw@ \advance \dimen@ -0\dp \tw@ \dimen@ii .08\wd \z@ \raise -\dimen@ \hbox to\dimen 4{\hss \kern \dimen@ii \box \tw@ \kern -\dimen@ii \hss
  }\protect \llap {\hbox to\dimen 4{\hss \box \z@ \hss }}}}{{\setbox \z@ \hbox {$\mathsurround \z@ \scriptscriptstyle D$}\setbox \tw@ \hbox {$\mathsurround \z@ \scriptscriptstyle /$}\dimen 4\wd \z@ \dimen@ \ht \tw@ \advance \dimen@ -\dp \tw@ \advance \dimen@ -\ht \z@ \advance \dimen@ \dp \z@ \divide \dimen@ \tw@ \advance \dimen@ -0\ht \tw@ \advance \dimen@ -0\dp \tw@ \dimen@ii .08\wd \z@ \raise -\dimen@ \hbox to\dimen 4{\hss \kern \dimen@ii \box \tw@ \kern -\dimen@ii \hss }\protect \llap {\hbox to\dimen 4{\hss \box \z@ \hss }}}}}_A)^{N_f}$ also includes a factor $m^{N_f n_0}$, where $n_0$ is the number of zero modes of ${\mathchoice {{\setbox \z@ \hbox {$\mathsurround \z@ \displaystyle D$}\setbox \tw@ \hbox {$\mathsurround \z@ \displaystyle /$}\dimen 4\wd \z@ \dimen@ \ht \tw@ \advance \dimen@ -\dp \tw@ \advance \dimen@ -\ht \z@ \advance \dimen@ \dp \z@ \divide \dimen@ \tw@ \advance \dimen@ -0\ht \tw@ \advance \dimen@ -0\dp \tw@ \dimen@ii .08\wd \z@ \raise -\dimen@ \hbox to\dimen 4{\hss \kern \dimen@ii \box
  \tw@ \kern -\dimen@ii \hss }\protect \llap {\hbox to\dimen 4{\hss \box \z@ \hss }}}}{{\setbox \z@ \hbox {$\mathsurround \z@ \textstyle D$}\setbox \tw@ \hbox {$\mathsurround \z@ \textstyle /$}\dimen 4\wd \z@ \dimen@ \ht \tw@ \advance \dimen@ -\dp \tw@ \advance \dimen@ -\ht \z@ \advance \dimen@ \dp \z@ \divide \dimen@ \tw@ \advance \dimen@ -0\ht \tw@ \advance \dimen@ -0\dp \tw@ \dimen@ii .08\wd \z@ \raise -\dimen@ \hbox to\dimen 4{\hss \kern \dimen@ii \box \tw@ \kern -\dimen@ii \hss }\protect \llap {\hbox to\dimen 4{\hss \box \z@ \hss }}}}{{\setbox \z@ \hbox {$\mathsurround \z@ \scriptstyle D$}\setbox \tw@ \hbox {$\mathsurround \z@ \scriptstyle /$}\dimen 4\wd \z@ \dimen@ \ht \tw@ \advance \dimen@ -\dp \tw@ \advance \dimen@ -\ht \z@ \advance \dimen@ \dp \z@ \divide \dimen@ \tw@ \advance \dimen@ -0\ht \tw@ \advance \dimen@ -0\dp \tw@ \dimen@ii .08\wd \z@ \raise -\dimen@ \hbox to\dimen 4{\hss \kern \dimen@ii \box \tw@ \kern -\dimen@ii \hss }\protect \llap {\hbox to\dimen 4{\hss \box \z@ \hss }}}}{{\setbox \z@
  \hbox {$\mathsurround \z@ \scriptscriptstyle D$}\setbox \tw@ \hbox {$\mathsurround \z@ \scriptscriptstyle /$}\dimen 4\wd \z@ \dimen@ \ht \tw@ \advance \dimen@ -\dp \tw@ \advance \dimen@ -\ht \z@ \advance \dimen@ \dp \z@ \divide \dimen@ \tw@ \advance \dimen@ -0\ht \tw@ \advance \dimen@ -0\dp \tw@ \dimen@ii .08\wd \z@ \raise -\dimen@ \hbox to\dimen 4{\hss \kern \dimen@ii \box \tw@ \kern -\dimen@ii \hss }\protect \llap {\hbox to\dimen 4{\hss \box \z@ \hss }}}}}_{A,0}$. Therefore an even integer $N_f$ ensures that $({\protect \rm det} {\mathchoice {{\setbox \z@ \hbox {$\mathsurround \z@ \displaystyle D$}\setbox \tw@ \hbox {$\mathsurround \z@ \displaystyle /$}\dimen 4\wd \z@ \dimen@ \ht \tw@ \advance \dimen@ -\dp \tw@ \advance \dimen@ -\ht \z@ \advance \dimen@ \dp \z@ \divide \dimen@ \tw@ \advance \dimen@ -0\ht \tw@ \advance \dimen@ -0\dp \tw@ \dimen@ii .08\wd \z@ \raise -\dimen@ \hbox to\dimen 4{\hss \kern \dimen@ii \box \tw@ \kern -\dimen@ii \hss }\protect \llap {\hbox to\dimen 4{\hss \box \z@ \hss
  }}}}{{\setbox \z@ \hbox {$\mathsurround \z@ \textstyle D$}\setbox \tw@ \hbox {$\mathsurround \z@ \textstyle /$}\dimen 4\wd \z@ \dimen@ \ht \tw@ \advance \dimen@ -\dp \tw@ \advance \dimen@ -\ht \z@ \advance \dimen@ \dp \z@ \divide \dimen@ \tw@ \advance \dimen@ -0\ht \tw@ \advance \dimen@ -0\dp \tw@ \dimen@ii .08\wd \z@ \raise -\dimen@ \hbox to\dimen 4{\hss \kern \dimen@ii \box \tw@ \kern -\dimen@ii \hss }\protect \llap {\hbox to\dimen 4{\hss \box \z@ \hss }}}}{{\setbox \z@ \hbox {$\mathsurround \z@ \scriptstyle D$}\setbox \tw@ \hbox {$\mathsurround \z@ \scriptstyle /$}\dimen 4\wd \z@ \dimen@ \ht \tw@ \advance \dimen@ -\dp \tw@ \advance \dimen@ -\ht \z@ \advance \dimen@ \dp \z@ \divide \dimen@ \tw@ \advance \dimen@ -0\ht \tw@ \advance \dimen@ -0\dp \tw@ \dimen@ii .08\wd \z@ \raise -\dimen@ \hbox to\dimen 4{\hss \kern \dimen@ii \box \tw@ \kern -\dimen@ii \hss }\protect \llap {\hbox to\dimen 4{\hss \box \z@ \hss }}}}{{\setbox \z@ \hbox {$\mathsurround \z@ \scriptscriptstyle D$}\setbox \tw@ \hbox {$\mathsurround
  \z@ \scriptscriptstyle /$}\dimen 4\wd \z@ \dimen@ \ht \tw@ \advance \dimen@ -\dp \tw@ \advance \dimen@ -\ht \z@ \advance \dimen@ \dp \z@ \divide \dimen@ \tw@ \advance \dimen@ -0\ht \tw@ \advance \dimen@ -0\dp \tw@ \dimen@ii .08\wd \z@ \raise -\dimen@ \hbox to\dimen 4{\hss \kern \dimen@ii \box \tw@ \kern -\dimen@ii \hss }\protect \llap {\hbox to\dimen 4{\hss \box \z@ \hss }}}}}_A)^{N_f}$ being positive for any nonzero $m$.}\BibitemShut {Stop}%
\bibitem [{\citenamefont {Witten}(1983)}]{witteninequality}%
  \BibitemOpen
  \bibfield  {author} {\bibinfo {author} {\bibfnamefont {E.}~\bibnamefont {Witten}},\ }\bibfield  {title} {\bibinfo {title} {Some inequalities among hadron masses},\ }\href {https://doi.org/10.1103/PhysRevLett.51.2351} {\bibfield  {journal} {\bibinfo  {journal} {Phys. Rev. Lett.}\ }\textbf {\bibinfo {volume} {51}},\ \bibinfo {pages} {2351} (\bibinfo {year} {1983})}\BibitemShut {NoStop}%
\bibitem [{\citenamefont {Vafa}\ and\ \citenamefont {Witten}(1984{\natexlab{a}})}]{vafawitten}%
  \BibitemOpen
  \bibfield  {author} {\bibinfo {author} {\bibfnamefont {C.}~\bibnamefont {Vafa}}\ and\ \bibinfo {author} {\bibfnamefont {E.}~\bibnamefont {Witten}},\ }\bibfield  {title} {\bibinfo {title} {Restrictions on symmetry breaking in vector-like gauge theories},\ }\href {https://doi.org/https://doi.org/10.1016/0550-3213(84)90230-X} {\bibfield  {journal} {\bibinfo  {journal} {Nuclear Physics B}\ }\textbf {\bibinfo {volume} {234}},\ \bibinfo {pages} {173} (\bibinfo {year} {1984}{\natexlab{a}})}\BibitemShut {NoStop}%
\bibitem [{\citenamefont {Vafa}\ and\ \citenamefont {Witten}(1984{\natexlab{b}})}]{vafawitten2}%
  \BibitemOpen
  \bibfield  {author} {\bibinfo {author} {\bibfnamefont {C.}~\bibnamefont {Vafa}}\ and\ \bibinfo {author} {\bibfnamefont {E.}~\bibnamefont {Witten}},\ }\bibfield  {title} {\bibinfo {title} {Eigenvalue inequalities for fermions in gauge theories},\ }\href {https://doi.org/10.1007/BF01212397} {\bibfield  {journal} {\bibinfo  {journal} {Communications in Mathematical Physics}\ }\textbf {\bibinfo {volume} {95}},\ \bibinfo {pages} {257} (\bibinfo {year} {1984}{\natexlab{b}})}\BibitemShut {NoStop}%
\bibitem [{\citenamefont {Kogan}\ \emph {et~al.}(1998)\citenamefont {Kogan}, \citenamefont {Kovner},\ and\ \citenamefont {Shifman}}]{Kogan}%
  \BibitemOpen
  \bibfield  {author} {\bibinfo {author} {\bibfnamefont {I.~I.}\ \bibnamefont {Kogan}}, \bibinfo {author} {\bibfnamefont {A.}~\bibnamefont {Kovner}},\ and\ \bibinfo {author} {\bibfnamefont {M.}~\bibnamefont {Shifman}},\ }\bibfield  {title} {\bibinfo {title} {Chiral symmetry breaking without bilinear condensates, unbroken axial ${Z}_{N}$ symmetry, and exact qcd inequalities},\ }\href {https://doi.org/10.1103/PhysRevD.59.016001} {\bibfield  {journal} {\bibinfo  {journal} {Phys. Rev. D}\ }\textbf {\bibinfo {volume} {59}},\ \bibinfo {pages} {016001} (\bibinfo {year} {1998})}\BibitemShut {NoStop}%
\bibitem [{\citenamefont {Golterman}(2024)}]{golterman}%
  \BibitemOpen
  \bibfield  {author} {\bibinfo {author} {\bibfnamefont {M.}~\bibnamefont {Golterman}},\ }\href {https://arxiv.org/abs/2406.02906} {\bibinfo {title} {Staggered fermions}} (\bibinfo {year} {2024}),\ \Eprint {https://arxiv.org/abs/2406.02906} {arXiv:2406.02906 [hep-lat]} \BibitemShut {NoStop}%
\bibitem [{\citenamefont {Chen}\ \emph {et~al.}(2014)\citenamefont {Chen}, \citenamefont {Lu},\ and\ \citenamefont {Vishwanath}}]{chenluashvin}%
  \BibitemOpen
  \bibfield  {author} {\bibinfo {author} {\bibfnamefont {X.}~\bibnamefont {Chen}}, \bibinfo {author} {\bibfnamefont {Y.-M.}\ \bibnamefont {Lu}},\ and\ \bibinfo {author} {\bibfnamefont {A.}~\bibnamefont {Vishwanath}},\ }\href@noop {} {\bibfield  {journal} {\bibinfo  {journal} {Nature Communications}\ }\textbf {\bibinfo {volume} {5}},\ \bibinfo {pages} {3507} (\bibinfo {year} {2014})}\BibitemShut {NoStop}%
\bibitem [{\citenamefont {Vishwanath}\ and\ \citenamefont {Senthil}(2013)}]{senthilashvin}%
  \BibitemOpen
  \bibfield  {author} {\bibinfo {author} {\bibfnamefont {A.}~\bibnamefont {Vishwanath}}\ and\ \bibinfo {author} {\bibfnamefont {T.}~\bibnamefont {Senthil}},\ }\bibfield  {title} {\bibinfo {title} {Physics of three-dimensional bosonic topological insulators: Surface-deconfined criticality and quantized magnetoelectric effect},\ }\href {https://doi.org/10.1103/PhysRevX.3.011016} {\bibfield  {journal} {\bibinfo  {journal} {Phys. Rev. X}\ }\textbf {\bibinfo {volume} {3}},\ \bibinfo {pages} {011016} (\bibinfo {year} {2013})}\BibitemShut {NoStop}%
\bibitem [{\citenamefont {Else}\ and\ \citenamefont {Thorngren}(2019)}]{else}%
  \BibitemOpen
  \bibfield  {author} {\bibinfo {author} {\bibfnamefont {D.~V.}\ \bibnamefont {Else}}\ and\ \bibinfo {author} {\bibfnamefont {R.}~\bibnamefont {Thorngren}},\ }\bibfield  {title} {\bibinfo {title} {Crystalline topological phases as defect networks},\ }\href {https://doi.org/10.1103/PhysRevB.99.115116} {\bibfield  {journal} {\bibinfo  {journal} {Phys. Rev. B}\ }\textbf {\bibinfo {volume} {99}},\ \bibinfo {pages} {115116} (\bibinfo {year} {2019})}\BibitemShut {NoStop}%
\bibitem [{\citenamefont {Li}\ \emph {et~al.}(2024)\citenamefont {Li}, \citenamefont {Oshikawa},\ and\ \citenamefont {Zheng}}]{Li_2024}%
  \BibitemOpen
  \bibfield  {author} {\bibinfo {author} {\bibfnamefont {L.}~\bibnamefont {Li}}, \bibinfo {author} {\bibfnamefont {M.}~\bibnamefont {Oshikawa}},\ and\ \bibinfo {author} {\bibfnamefont {Y.}~\bibnamefont {Zheng}},\ }\bibfield  {title} {\bibinfo {title} {Decorated defect construction of gapless-spt states},\ }\bibfield  {journal} {\bibinfo  {journal} {SciPost Physics}\ }\textbf {\bibinfo {volume} {17}},\ \href {https://doi.org/10.21468/scipostphys.17.1.013} {10.21468/scipostphys.17.1.013} (\bibinfo {year} {2024})\BibitemShut {NoStop}%
\bibitem [{\citenamefont {Ma}\ and\ \citenamefont {Wang}(2023)}]{mawang}%
  \BibitemOpen
  \bibfield  {author} {\bibinfo {author} {\bibfnamefont {R.}~\bibnamefont {Ma}}\ and\ \bibinfo {author} {\bibfnamefont {C.}~\bibnamefont {Wang}},\ }\bibfield  {title} {\bibinfo {title} {Average symmetry-protected topological phases},\ }\href {https://doi.org/10.1103/PhysRevX.13.031016} {\bibfield  {journal} {\bibinfo  {journal} {Phys. Rev. X}\ }\textbf {\bibinfo {volume} {13}},\ \bibinfo {pages} {031016} (\bibinfo {year} {2023})}\BibitemShut {NoStop}%
\bibitem [{\citenamefont {Witten}(1984)}]{Witten1984}%
  \BibitemOpen
  \bibfield  {author} {\bibinfo {author} {\bibfnamefont {E.}~\bibnamefont {Witten}},\ }\bibfield  {title} {\bibinfo {title} {Non-abelian bosonization in two dimensions},\ }\href {https://doi.org/10.1007/BF01215276} {\bibfield  {journal} {\bibinfo  {journal} {Communications in Mathematical Physics}\ }\textbf {\bibinfo {volume} {92}},\ \bibinfo {pages} {455} (\bibinfo {year} {1984})}\BibitemShut {NoStop}%
\bibitem [{\citenamefont {lan Affleck}(1986)}]{affleck1986}%
  \BibitemOpen
  \bibfield  {author} {\bibinfo {author} {\bibnamefont {lan Affleck}},\ }\href@noop {} {\bibfield  {journal} {\bibinfo  {journal} {Nucl. Phys. B}\ }\textbf {\bibinfo {volume} {265}},\ \bibinfo {pages} {409} (\bibinfo {year} {1986})}\BibitemShut {NoStop}%
\bibitem [{\citenamefont {James}\ \emph {et~al.}(2018)\citenamefont {James}, \citenamefont {Konik}, \citenamefont {Lecheminant}, \citenamefont {Robinson},\ and\ \citenamefont {Tsvelik}}]{James_2018}%
  \BibitemOpen
  \bibfield  {author} {\bibinfo {author} {\bibfnamefont {A.~J.~A.}\ \bibnamefont {James}}, \bibinfo {author} {\bibfnamefont {R.~M.}\ \bibnamefont {Konik}}, \bibinfo {author} {\bibfnamefont {P.}~\bibnamefont {Lecheminant}}, \bibinfo {author} {\bibfnamefont {N.~J.}\ \bibnamefont {Robinson}},\ and\ \bibinfo {author} {\bibfnamefont {A.~M.}\ \bibnamefont {Tsvelik}},\ }\bibfield  {title} {\bibinfo {title} {Non-perturbative methodologies for low-dimensional strongly-correlated systems: From non-abelian bosonization to truncated spectrum methods},\ }\href {https://doi.org/10.1088/1361-6633/aa91ea} {\bibfield  {journal} {\bibinfo  {journal} {Reports on Progress in Physics}\ }\textbf {\bibinfo {volume} {81}},\ \bibinfo {pages} {046002} (\bibinfo {year} {2018})}\BibitemShut {NoStop}%
\bibitem [{\citenamefont {Delmastro}\ \emph {et~al.}(2023)\citenamefont {Delmastro}, \citenamefont {Gomis},\ and\ \citenamefont {Yu}}]{delmastro2023}%
  \BibitemOpen
  \bibfield  {author} {\bibinfo {author} {\bibfnamefont {D.}~\bibnamefont {Delmastro}}, \bibinfo {author} {\bibfnamefont {J.}~\bibnamefont {Gomis}},\ and\ \bibinfo {author} {\bibfnamefont {M.}~\bibnamefont {Yu}},\ }\href {https://arxiv.org/abs/2108.02202} {\bibinfo {title} {Infrared phases of 2d qcd}} (\bibinfo {year} {2023}),\ \Eprint {https://arxiv.org/abs/2108.02202} {arXiv:2108.02202 [hep-th]} \BibitemShut {NoStop}%
\bibitem [{\citenamefont {Shifman}\ and\ \citenamefont {Yung}(2004)}]{susyvortex1}%
  \BibitemOpen
  \bibfield  {author} {\bibinfo {author} {\bibfnamefont {M.}~\bibnamefont {Shifman}}\ and\ \bibinfo {author} {\bibfnamefont {A.}~\bibnamefont {Yung}},\ }\bibfield  {title} {\bibinfo {title} {Non-abelian string junctions as confined monopoles},\ }\href {https://doi.org/10.1103/PhysRevD.70.045004} {\bibfield  {journal} {\bibinfo  {journal} {Phys. Rev. D}\ }\textbf {\bibinfo {volume} {70}},\ \bibinfo {pages} {045004} (\bibinfo {year} {2004})}\BibitemShut {NoStop}%
\bibitem [{\citenamefont {Hanany}\ and\ \citenamefont {Tong}(2003)}]{susyvortex2}%
  \BibitemOpen
  \bibfield  {author} {\bibinfo {author} {\bibfnamefont {A.}~\bibnamefont {Hanany}}\ and\ \bibinfo {author} {\bibfnamefont {D.}~\bibnamefont {Tong}},\ }\bibfield  {title} {\bibinfo {title} {Vortices, instantons and branes},\ }\href {https://doi.org/10.1088/1126-6708/2003/07/037} {\bibfield  {journal} {\bibinfo  {journal} {Journal of High Energy Physics}\ }\textbf {\bibinfo {volume} {2003}},\ \bibinfo {pages} {037–037} (\bibinfo {year} {2003})}\BibitemShut {NoStop}%
\bibitem [{\citenamefont {Tong}(2009)}]{susyvortex3}%
  \BibitemOpen
  \bibfield  {author} {\bibinfo {author} {\bibfnamefont {D.}~\bibnamefont {Tong}},\ }\bibfield  {title} {\bibinfo {title} {Quantum vortex strings: A review},\ }\href {https://doi.org/10.1016/j.aop.2008.10.005} {\bibfield  {journal} {\bibinfo  {journal} {Annals of Physics}\ }\textbf {\bibinfo {volume} {324}},\ \bibinfo {pages} {30–52} (\bibinfo {year} {2009})}\BibitemShut {NoStop}%
\end{thebibliography}%

\appendix

\end{document}